\documentclass{emulateapj}

\newcommand{\NuSTAR}{\textit{NuSTAR}}
\newcommand{\XMM}{\textit{XMM-Newton}}
\newcommand{\Chandra}{\textit{Chandra}}

\usepackage{graphicx}
\usepackage{bm}
\usepackage{color}
\usepackage{booktabs}
\usepackage{gensymb}

\begin{document}

\title{Hard X-ray View of HCG 16 (Arp 318)}

\author{Saeko Oda\altaffilmark{1},
Yoshihiro Ueda\altaffilmark{1},
Atsushi Tanimoto\altaffilmark{1},
Claudio Ricci\altaffilmark{2,3,4}}

\affil{
\altaffilmark{1}{Department of Astronomy, Kyoto University, Kyoto 606-8502, Japan \\}
\altaffilmark{2}{Instituto de Astrofisica, Pontificia Universidad Catolica de Chile, Casilla 306, Santiago 22, Chile \\}
\altaffilmark{3}{Kavli Institute for Astronomy and Astrophysics, Peking University, Beijing 100871, China \\}
\altaffilmark{4}{Chinese Academy of Sciences South America Center for Astronomy and
China-Chile Joint Center for Astronomy, Camino El Observatorio 1515, Las Condes, Santiago, Chile}}

\begin{abstract}

We report the hard X-ray (3--50 keV) view of the compact group HCG 16
(Arp 318) observed with \textit{Nuclear Spectroscopic Telescope Array}
(\NuSTAR). NGC~838 and NGC~839 are undetected at energies above 8 keV, 
showing no evidence of heavily obscured active galactic nuclei (AGNs).
This confirms that these are starburst-dominant galaxies as previously
suggested. We perform a comprehensive broadband (0.3--50 keV) X-ray
spectral analysis of the interacting galaxies NGC~833 and NGC~835, using
data of \NuSTAR, \Chandra, and \XMM \  observed on multiple epochs
from 2000 to 2015. \NuSTAR\ detects the transmitted continua of
low-luminosity active galactic nuclei (LLAGNs) in NGC~833 and NGC~835
with line-of-sight column densities of $\approx
3\times10^{23}$~cm$^{-2}$ and intrinsic 2--10 keV luminosities of
$\approx 3\times10^{41}$ erg s$^{-1}$.  
The iron-K$\alpha$ to hard X-ray luminosity ratios of NGC~833 and NGC~835
suggest that their tori are moderately developed, which may have been
triggered by the galaxy interactions. We find that
NGC~835 underwent long-term variability in both
intrinsic luminosity (by a factor of 5) and absorption (by $\Delta
N_{\rm H} \approx 2\times10^{23}$ cm$^{-2}$). We discuss
the relation between the X-ray and total infrared luminosities in local
LLAGNs hosted by spiral galaxies. The large diversity in their ratios
is consistent with the the general idea
that the mass accretion process in the nucleus and the star forming
activity in the disk are not strongly coupled, regardless of the 
galaxy environment.

\end{abstract}

\keywords{galaxies: active -- galaxies: individual (HCG~16) -- X-rays: galaxies}

\section{Introduction}

It is generally believed that galactic bulges and the supermassive black holes
(SMBHs) in their centers ``co-evolve'' (see \citealt{kor13} for a
review), although a detailed understanding of this process is still missing.
One of the mechanisms responsible for the co-evolution
could be major mergers, which cause
vigorous starburst activity together with rapid mass accretion onto
SMBHs \citep[e.g.,][]{hop08}.
During this accreting phases, systems would be observed
as an Active Galactic Nuclei (AGNs), and emit strongly in the X-ray band.
In the late stages of the merger process, the nuclei are expected to be heavily
obscured by gas and dust, and even become Compton-thick,
with hydrogen column densities of $N_\mathrm{H} \geq 10^{24}$ cm$^{-2}$.
To detect such hidden AGNs in merging galaxies,
hard X-ray observations at energies above 10 keV are a promising
approach, being the least biased against
heavy obscuration. In fact, recent studies using
\textit{Nuclear Spectroscopic Telescope Array} (\NuSTAR; \citealt{har13}), which achieve
the best sensitivity above 10 keV to date, have revealed the presence of
heavily obscured AGNs in many interacting galaxies in the local Universe
(e.g., \citealt{ricci17}).

\begin{figure}
\begin{center}
\plotone{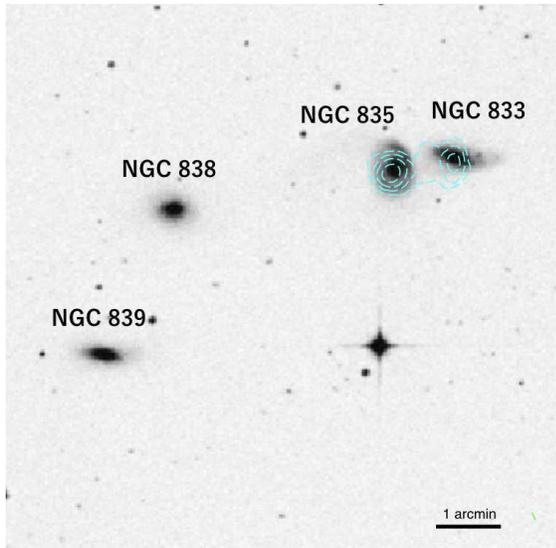}
\end{center}
\caption{
Digitized Sky Survey 2 (DSS2) R-band image of the four major galaxies in
HCG~16. 
The image size is 8.5 arcmin square. 
Hard X-ray contours of the \NuSTAR\ 
3--24 keV image are overlaid with the cyan dotted lines.}
\label{dss}
\end{figure}

HCG~16 \citep{hic82}, also known as Arp~318, is one of the nearest
compact groups of galaxies and hence an ideal target to study
galaxy interactions and its influences on AGN activities.  It
consists of seven member galaxies, which include the central four spiral
galaxies originally identified by \citet{hic82}, NGC~833, NGC~835,
NGC~838, and NGC~839. 
\citet{ver01} and \citet{gal08} determined that
HGC~16 is an intermediate evolutionary stage of compact groups.
Table~\ref{info} summarizes the basic properties of the four major
galaxies in HCG~16. 
Figure~\ref{dss} displays the Digitized Sky Survey
R-band image of the central region of HCG~16. 
NGC~833 is lopsided because of a recent interaction with
NGC~835, which has a east tidal tail toward NGC~838 (\citealt{kon13}).
\citet{ver10} optically classified NGC 833 as a LINER, and NGC 835 as a
Seyfert 2. NGC~838 is included in the sample of the Great Observatories
All-sky LIRG Survey (GOALS; \citealt{arm09}),
and is classified as a starburst 
galaxy (\citealt{san03}; \citealt{bit14}). NGC~839 is optically
classified as a LINER. 
These four galaxies are luminous in the mid-to-far infrared band,  
indicative of powerful energy sources (starburst and/or AGN), 
and have been investigated with \textit{Spitzer} and \textit{Herschel}.
Following an earlier work by \citet{gal08}, who
systematically analyzed 
the \textit{Spitzer} photometric data of 12 nearby HCGs,
\citet{bit14} estimated
the star formation rates and the 8--1000 $\mu$m infrared luminosities
of the individual galaxies in HCG~16 by fitting 
the UV to submillimeter
spectral energy distributions (SEDs) obtained 
with GALEX, SDSS, \textit{Spitzer}, and \textit{Herschel}.
These luminosities are summarized in Table~\ref{info}.

\begin{deluxetable*}{cccccccc}
\tablecaption{Basic Information of 4 Major Member Galaxies in HCG 16 \label{info}}
\tablewidth{\textwidth}
\tablehead{  Name & HCG Name & Morphology & Nuclear classification & $\alpha$  & $\delta$ & SFR & log $L_\mathrm{IR}$ \\
  &  & & & (J2000) & (J2000) & $[M_\odot$ yr$^{-1}$] & [$L_\odot$] \\
   &  & (1) & (2) & (3) & (3) & (4) & (5) } 
\startdata
NGC~833 & HCG~16B & Sab   & LINER &  02$^\mathrm{h}$09$^\mathrm{m}$20.8$^\mathrm{s}$  
& -10$\degree$07$'$59$''$   &  0.04  &  9.74\\
NGC~835 & HCG~16A & SBab  & Seyfert 2 & 02$^\mathrm{h}$09$^\mathrm{m}$24.6$^\mathrm{s}$  
& -10$\degree$08$'$09$''$  &  3.10  & 10.76\\
NGC~838 & HCG~16C & Im     & Starburst &  02$^\mathrm{h}$09$^\mathrm{m}$38.5$^\mathrm{s}$  
& -10$\degree$08$'$48$''$   &  11.5  & 11.09    \\
NGC~839 & HCG~16D & Im     & LINER & 02$^\mathrm{h}$09$^\mathrm{m}$42.9$^\mathrm{s}$  
& -10$\degree$11$'$03$''$   &  1.16  & 10.94         
\enddata
\tablecomments{
(1) The galaxy morphology based on \citet{hic89}.
(2) The classification of the nucleus based on \citet{gal08}.
(3) The coordinates (R.A. and Dec.). 
(4) The star formation rates estimated by \citet{bit14}.
(5) The 8--1000 $\mu$m infrared luminosities based on the \textit{Herschel} observations (\citealt{bit14}).
}
\end{deluxetable*}

X-ray data provide unique insights on the dominant 
energy sources of infrared galaxies and the properties of their AGNs.  
The X-ray emission from the HGC~16 system was first detected by the \textit{Einstein}
Observatory (\citealt{bah84}). 
Using 
\textit{X-ray Multi-Mirror Mission} (\XMM, \citealt{jan01}), \citet{tur01_hcg16} 
discovered obscured AGNs in NGC~833, NGC~835, and NGC~839, 
which co-exist with starburst activities. They found no evidence for AGN
in NGC~838, and hence classified it as a pure starburst galaxy. 
\Chandra \ observations of HCG~16 have been conducted five times 
between 2000 and 2013 (see Table~\ref{obs}), including the first
one reported by \citet{gon06}. 
Analyzing the spectra of all the five \Chandra\ observations,
\citet{sul14} confirmed the presence of obscured AGNs in NGC~833 and
NGC~835, whereas they concluded that NGC~838 and NGC~839 are
starburst-dominated galaxies.
Although the presence of a heavily
obscured and/or faint AGN in NGC~839 was not excluded, they suggested
that its hard X-ray emission is likely to arise from high
mass X-ray binaries (HMXBs). Also, \citet{sul14} found that the X-ray
flux of NGC~835 increased compared with the 2000 and 2008
observations, mainly due to changes in the luminosity.
Focusing on the long-term time variability of NGC~835, \citet{gon16} presented more
detailed studies with \Chandra. They compared the X-ray luminosities
with the mid-infrared ones 
obtained from the high spatial resolution CanariCam/GTC data 
and concluded
that the X-ray variability is due to a change in the
column density, and not to variability of the intrinsic flux.

Thus, there still remain unsettled issues on X-ray properties of the
member galaxies in HCG~16. Sensitive hard X-ray data above $\sim$8 keV
are useful to tackle these problems, and allow us to determine the AGN
intrinsic luminosity and absorption with the best accuracy.  In this
paper, we report the results of the first hard X-ray ($>10$ keV) imaging
observation of HCG~16 performed with \NuSTAR \  in 2015, 
which covered the four major galaxies. 
We particularly focus on the
broadband X-ray spectral analysis of NGC~833 and NGC~835 including 
the \Chandra \ and \XMM\ data to best understand the
nature of their AGNs.
The paper is organized as follows. In Section~2,
we present the observations and data reduction.
Section~3 describes the details of the broadband X-ray spectral analysis 
of NGC~833 and NGC~835. Section~4 presents new constraints from the \NuSTAR\ 
observation on the nature of NGC~838 and NGC~839.
We discuss the implications of our results in Section~5 and 
summarize our findings in Section~6.
Throughout the paper, 
we adopt a distance of 34~Mpc \citep{gon16} for all the member galaxies,
which corresponds to a redshift of $z$=0.007939 for the cosmological
parameters $H_0 = 70 $ km s$^{-1}$ Mpc$^{-1}$, $\Omega_\mathrm{m} =
0.27$, and $\Omega_\lambda = 0.73$. 
We assume the solar
abundances by \citet{and89} and the photoelectric absorption
cross-sections by \citet{bal92}. All the uncertainties in spectral
parameters correspond to the 90\% confidence level for a single
parameter of interest.

\section{Observations and Data Reduction}

Table~\ref{obs} gives the log of of all the \NuSTAR, \Chandra, and \XMM
\ observations of HCG 16 performed up to date. The details of the
observations and the data reduction procedures of each satellite are
described below. 
We analyze the \NuSTAR, \Chandra, and \XMM\ data of NGC 833 and NGC 835
to perform simultaneous spectral fitting (Section~3). Since NGC 838 and
NGC 839 are not detected by \NuSTAR\ (Section~2.1), we do not analyze
the \Chandra\ and \XMM\ data of these galaxies, whose results were
already reported in detail by \citet{sul14} and \citet{tur01_hcg16}.

\begin{deluxetable*}{cccccc}
\tablecaption{Observation Log of HCG 16\label{obs}}
\tablewidth{\textwidth}
\tablehead{  & Instrument  &  Observation ID  &  Start Time  &  End Time  &  Exposure  \\
  &  &  & [UT] & [UT] & [ks] \\
  &  &  & (1) & (1) & (2) } 
\startdata
 \NuSTAR  &   FPMA, FPMB  &  60061346002 &  2015 September 13 05:11 &  2015 September 13 15:56 & 18.1/18.4  \\
\Chandra  &  ACIS  & 15667 & 2013 July 21 10:58 & 2013 July 22 04:02 & 58.3\\
  & & 15666 & 2013 July 18 21:06 & 2013 July 19 06:15 & 29.7 \\
  & & 15181 & 2013 July 16 04:49 & 2013 July 16 19:28 & 49.5 \\
 & & 923  & 2000 November 16 21:47 &  2000 November 01:58 & 12.6  \\
\XMM  &  EPIC-pn, EPIC-MOS1,2  & 0115810301  & 2000 January 23 17:30 & 2000 January 24 08:52 &  34.7/46.2/43.1 
\enddata
\tablecomments{
(1)~Based on FPMA for \NuSTAR \  and EPIC-MOS1 for \XMM. 
(2)~The exposure time of \NuSTAR \ (FPMA/FPMA), \Chandra,
and \XMM \ (EPIC-pn/EPIC-MOS1/EPIC-MOS2).
}
\end{deluxetable*}

\subsection{NuSTAR}

\begin{figure}
\begin{center}
\plotone{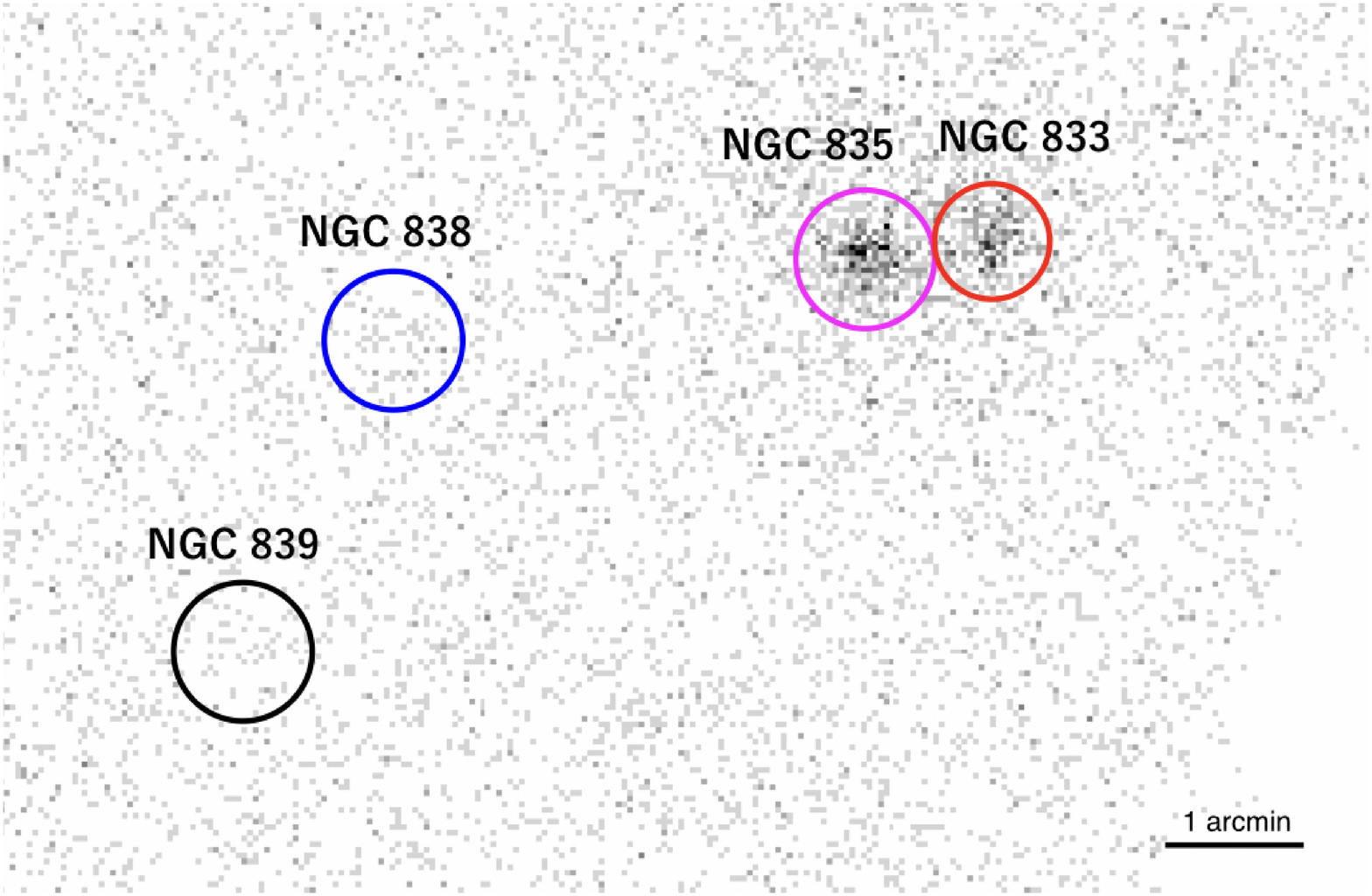}
\end{center}
\caption{\NuSTAR\ image of HCG~16 the 3--79 keV band. The spectral
extraction regions are marked with red, magenta, blue, and black
circles for NGC~833 ($25''$ radius), NGC~835 ($30''$), NGC~838 ($30''$),
and NGC~839 ($30''$), respectively.}
\label{nustar}
\end{figure}

The \NuSTAR \ mission, launched on 2012 June 13, is the first
astronomical observatory that employs focusing optics in the hard X-ray
band above 10 keV. It carries two co-aligned grazing
incidence telescopes coupled with two focal plane modules (FPMs):
FPMA and FPMB, and covers the 3--79 keV band. HCG~16 was observed by
\NuSTAR\ on 2015 September 13 for a net exposure of 36.6~ks. We reduced
the data using HEAsoft version 6.19 and calibration database (CALDB)
version 2016 September 22. We created calibrated event files with the
\textit{nupipeline} script after removing any background
flares. Figure~\ref{nustar} displays the combined image of FPMA and FPMB
in the 3-79 keV band.

Strong signals were detected from NGC~833 and NGC~835, for which we
perform detailed broadband spectral analysis combining the \Chandra \  and
\XMM \ data in Section~3.
We used the \textit{nuproducts} script for spectral extraction. The
angular separation between NGC~833 and NGC~835 is $55.9''$, which is
large enough to reduce the individual spectra of the two targets with
\NuSTAR . We selected source photon events from circular regions with
radii of $25''$ and $30''$ for NGC~833 and NGC~835, respectively. The
background (cosmic X-ray background plus non X-ray background) was taken
from a source-free circular region with a radius of $50''$, and was
used for all the targets. We confirmed that the flux
normalizations of FPMA and FPMB spectra were consistent with each
other. Thus, we co-added the spectra of FPMA and FPMB for each
target. 
Although NGC~833 and NGC~835 are spatially well resolved
by \NuSTAR \ and their spectra are mostly separated,
small contamination to one spectrum from the
other is expected because of the tail of the point spread function of
\NuSTAR\ beyond $>1'$.
The fractions of this contamination 
are estimated to be $\sim 11$\% and $\sim 10$\%
in the total fluxes (not including the background)
contained in the NGC 833 and NGC 835 spectrum-extraction regions,
respectively, by performing spectral simulations.
We subtracted it as an additional background for each target.

\NuSTAR \ failed to detect significant signals of NGC~838 and NGC~839.
To evaluate the upper limits, we extracted the source spectra of these two
galaxies from circular regions with a radius of $30''$. FPMA and FPMB
spectra were combined using the script \textit{addascaspec}. We find that
3$\sigma$ upper limits of the count rates in the 3--8 keV and 8--24 keV
bands are respectively 0.0011 and 0.00084 counts s$^{-1}$ for NGC~838,
0.00072 and 0.00061 counts s$^{-1}$ for NGC~839. These upper limits
are all larger than those predicted from the model adopted by
\citet{sul14}. We discuss this result in Section~4.

\begin{deluxetable*}{ccccccc}
\tablecaption{Summary of Data Reduction of NGC 833\label{reduc1}}
\tablewidth{\textwidth}
\tablehead{  & Instrument  & Source radius & Background radius & Counts per energy bin & Energy band & Net count rate   \\
  &  & [arcsec] & [arcsec] & [counts] & [keV] & [$10^{-2}$ \ count \ s$^{-1}$]\\
   &  & (1) & (2) & (3) & (4) & (5) } 
\startdata
 \NuSTAR  &  FPMs & 25 & 50 & 50 & 4.5--40 & 1.13 \\
 \Chandra & ACIS (2013) & 15 & 20 & 50 & 0.3--9 & 0.94 \\
   & ACIS (2000) & 15 & 20 & 25 &
   0.75--8 & 1.93 \\
\XMM & EPIC-pn & 25 & 25 & 20 & 0.5--8.5 & 0.67 \\
  & EPIC-MOSs & 25 &25 & 50 & 0.45--10 & 1.26 
\enddata
\tablecomments{
(1) The radius of the spectral extraction region for the source.
(2) The radius of the spectral extraction region for the background.
(3) The minimum photon counts of each spectral bin.
(4) The energy band used in the spectral analysis.
(5) The net count rate in this energy band after background subtraction.
}
\end{deluxetable*}

\begin{deluxetable*}{ccccccc}
\tablecaption{Summary of Data Reduction of NGC 835  \label{reduc2}}
\tablewidth{\textwidth}
\tablehead{  & Instrument  & Source radius & Background radius & Counts per energy bin & Energy band & Net count rate   \\
  &  & [arcsec] & [arcsec] & [counts] & [keV] & [$10^{-2}$ \ count \ s$^{-1}$]\\
   &  & (1) & (2) & (3) & (4) & (5) } 
\startdata
 \NuSTAR  &  FPMs & 30 & 50 & 50 & 4.5--49 & 1.82 \\
 \Chandra & ACIS (2013) & 15 & 20 & 50 & 0.5--9 & 3.18 \\
  & ACIS (2000) & 15 & 20 & 25 & 
  0.45--7 & 2.58 \\
\XMM  & EPIC-pn & 25 & 25 & 50 & 0.3--7.5 & 5.49 \\
& EPIC-MOSs & 25 &25 & 50 & 0.3--8 & 1.54 
\enddata
\tablecomments{
(1)--(5) The same parameters as those of Table \ref{reduc1}.}
\end{deluxetable*}

\subsection{Chandra}

HCG~16 was observed five times with the \Chandra \ X-ray observatory
(CXO; \citealt{wei02}), which carries two focal plane science
instruments: the High-Resolution Camera (HRC) and the Advanced CCD
Imaging Spectrometer (ACIS). The four major galaxies were covered by
the ACIS-S3 chip in ObsIDs 923, 15181, 15666 and 15667, whereas only
NGC~835 and NGC~838 were observed in ObsID 10394.  We reduced the data
following the standard guidelines,
using the \Chandra \ Interactive Analysis of
Observations (CIAO) version 4.8 and the latest CALDB version 4.7.2.

We utilized the script \textit{chandra\_repro} and \textit{specextract}
to produce the spectra of NGC~833 and NGC~835 for each ObsID. Checking
the background light curves, we confirmed that none of the data suffered
from background flares.
We did not use the ObsID 10394 data (2008) for our spectral fitting,
since NGC~833 was not in the field-of-view and 
NGC~835 was located close to the edge of the camera, with poor photon statistics available.
As no significant spectral variability was found among
the 3 observations carried out in July 2013, we co-added these spectra using the
\textit{addascaspec} task.

\begin{figure}[t!]
\begin{center}
\plotone{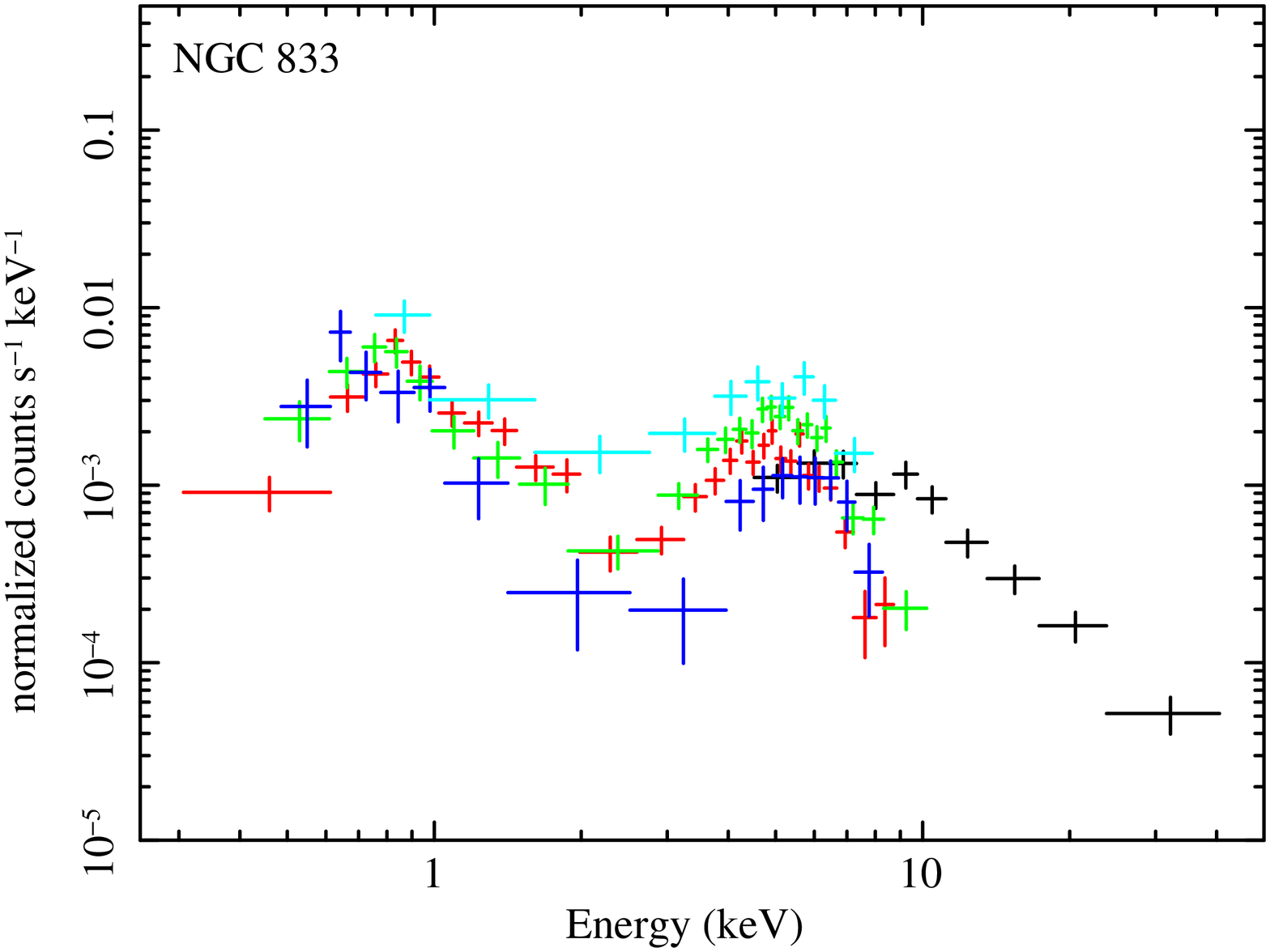}
\plotone{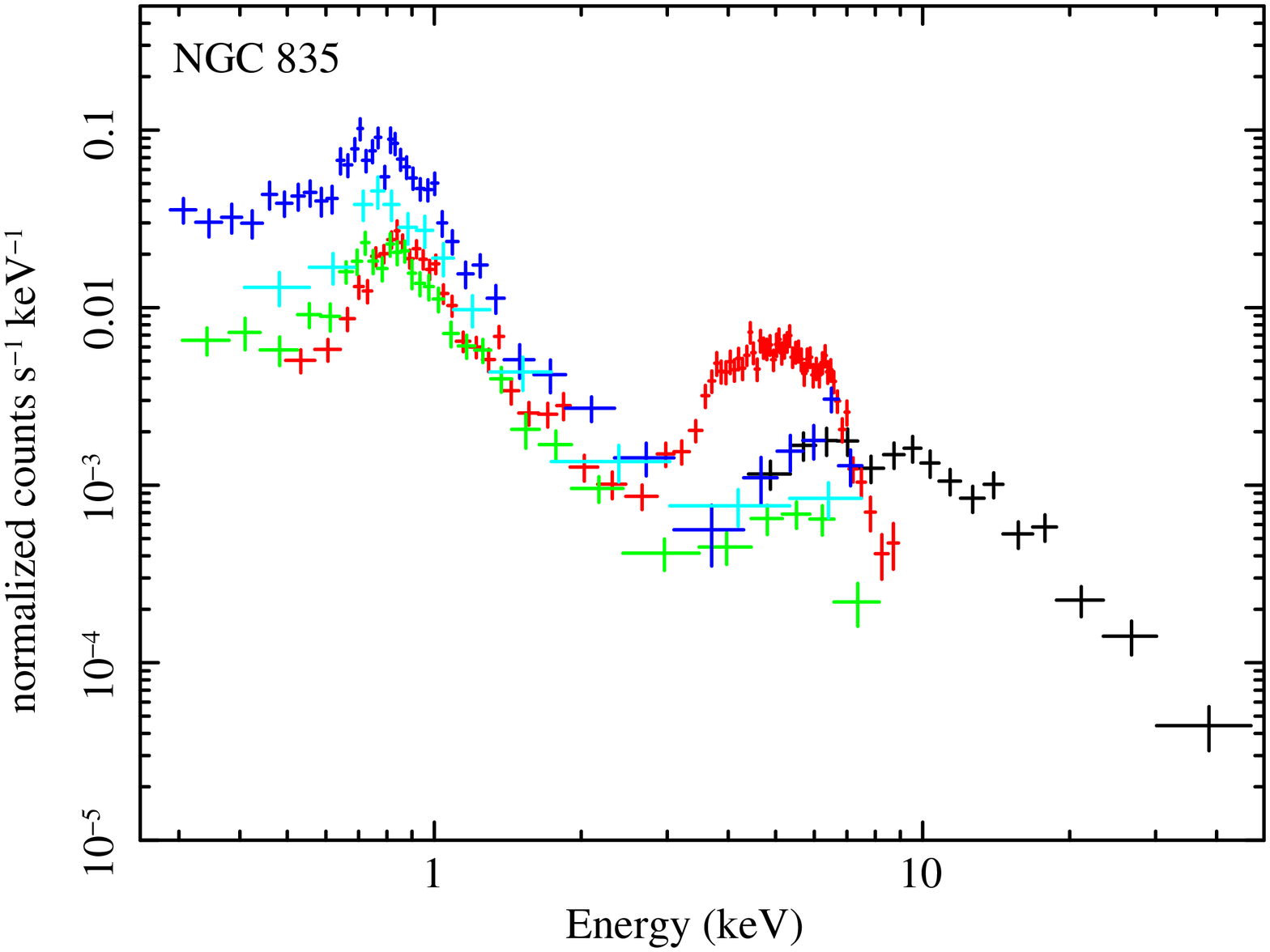}
\end{center}
\caption{Observed spectra of NGC~833 and NGC~835 folded with the energy
 responses. The black, red, cyan, blue, and green crosses indicate the data of FPMs, ACIS (2013),
 ACIS (2000), EPIC-pn, and EPIC-MOS, respectively.\\}
\label{rawdata}
\end{figure}

\subsection{XMM-Newton}

\XMM\ observed
HCG~16 on 2000 January 23 and 24
as a first-light observation\footnote{
We have confirmed that the first-light data can be validly reduced with 
the standard pipeline processing (Ehle, M., private communication).}
\XMM\ carries three X-ray CCD cameras
on board: one EPIC-pn (\citealt{str01}) and two EPIC-MOS
(\citealt{tur01_mos}). The data were reprocessed with the Science
Analysis Software (SAS) version 15.0.0 and using the Current Calibration File
(CCF) of 2016 May. The pipeline scripts \textit{epproc} for EPIC-pn and
\textit{emproc} for MOS were used to produce calibrated photon event
files. We selected only PATTERN $\le$ 4 and PATTERN $\le$ 12 events for
EPIC-pn and EPIC-MOS, respectively. We excluded data suffering from
background flares when the count rates above 10 keV exceeded 2~count
s$^{-1}$ for EPIC-pn and 0.4 count s$^{-1}$ for EPIC-MOS. The net
exposures are 34.7 ks (EPIC-pn), 46.2 ks (EPIC-MOS1), and 43.1 ks (EPIC-MOS2).

The source spectra of NGC~833 and NGC~835 were extracted from circular
regions with a radius of $25''$ for all the cameras. The background
spectrum was taken from a source-free region in the same CCD chip.  We
created the Redistribution Matrix File (RMF) and Ancillary Response File
(ARF) with the \textit{rmfgen} and \textit{arfgen} task,
respectively. The source spectra, background spectra, RMFs, and ARFs of
EPIC-MOS1 and EPIC-MOS2 were merged by using the \textit{addascaspec}
script.\\

\section{Broadband Spectral Analysis of NGC~833 and NGC~835}

We simultaneously analyze the broadband spectra of NGC~833 and NGC~835
observed with \NuSTAR/FPMs, \Chandra/ACIS, \XMM/EPIC-pn, and
EPIC-MOSs. We determine the energy band used for our analysis by
considering the signal-to-noise ratio of each spectrum. The detailed
information of the spectra (including binning and net count rate) is
summarized in Tables~\ref{reduc1} and \ref{reduc2} for NGC~833 and NGC~835, respectively.
XSPEC 12.9.0s is utilized for spectral fitting with $\chi^2$ statistics.

Figure~\ref{rawdata} plots the spectra of NGC~833 and NGC~835 folded
with the energy responses. To consider the Galactic absorption, we
multiply \textbf{phabs} to all models and fix the hydrogen column
density to $N_\mathrm{H}^\mathrm{Gal} = 2.37 \times 10^{20} \
\mathrm{cm}^{-2}$, as estimated by \citet{kal05}.
Cross calibration
uncertainties among different instruments are taken into account by multiplying a
constant factor (\textbf{const0}) with respect to the FPMs and ACIS
spectra\footnote{The cross-normalizations between the FPMs and ACIS 
transmission grating spectrosmeters
are consistent within $\sim$10\% according to \citet{mad15}.},
which are taken as the calibration references in this paper. 
The \textbf{const0} factor is left free for EPIC-pn and EPIC-MOSs.

\tabletypesize{\scriptsize}
\begin{deluxetable}{cccc}[t]
\tablecaption{Best-fit Parameters of NGC~833 \label{par1}}
\tablewidth{0.5\textwidth}
\tablehead{
 Note$^{*}$ & Parameter  & base model  &   e-torus model  
}
\startdata
(1) &  $ N_\mathrm{H}^\mathrm{LS} \  [10^{23} \ \mathrm{cm^{-2} }]$   & $2.68^{+0.37}_{-0.32}$ & $2.54^{+0.31}_{-0.25}$\\
(2) &  $\Gamma_\mathrm{AGN}$ &  $1.69^{+0.26}_{-0.25}$  & $1.57^{+0.23}_{-0.07}$ \\
(3) & $A_\mathrm{AGN} \ [10^{-4} \ \mathrm{keV^{-1}\ cm^{-2} \ s^{-1}}]$   & $5.2^{+4.7}_{-2.4}$ & $3.8^{+3.0}_{-1.4}$ \\
(4) &  $|R|$ & $0.31^{+0.31}_{-0.26}$ & -- \\
(5) & $f_\mathrm{scat}\ [\%] $   &  $1.01^{+0.96}_{-0.51}$  & $1.46^{+0.45}_{-0.70}$ \\
(6) & $\Gamma_\mathrm{scat}$ & $2.06^{+0.31}_{-0.35}$ & $2.01^{+0.37}_{-0.45}$  \\
(7) & $k \mathrm{T} \ [\mathrm{keV}] $   &  $0.60^{+0.10}_{-0.14}$  & $0.60^{+0.10}_{-0.15}$ \\  
(8) &  $ A_\mathrm{apec}\ [10^{-6} \ \mathrm{cm^{-5}}]$  &  $3.00^{+0.66}_{-0.64}$  & $2.93^{+0.76}_{-0.75}$ \\
(9) & $N_\mathrm{Ch13}^\mathrm{time}$  &  $0.47^{+0.09}_{-0.07}$  & $0.49^{+0.08}_{-0.07}$ \\
(10) & $N_\mathrm{Ch00}^\mathrm{time}$  &  $1.39^{+0.34}_{-0.28}$  & $1.44^{+0.34}_{-0.29}$ \\
(11) & $N_\mathrm{XMM}^\mathrm{time}$  & $0.73^{+0.17}_{-0.14}$ & $0.75^{+0.13}_{-0.14}$ \\
(12) & $N_\mathrm{MOS}$   &  $1.22^{+0.17}_{-0.16}$  &  $1.22^{+0.17}_{-0.15}$ \\
(13) &  $N_\mathrm{pn}$  & $1.02^{+0.20}_{-0.18}$  & $1.02^{+0.20}_{-0.18}$ \\
(14) &  $F_{2-10} \ [\mathrm{erg \ cm^{-2} \ s^{-1}}]$  & $ 6.3 \times 10^{-13}$ & $ 6.2 \times 10^{-13}$ \\
(15) &  $F_{10-50} \ [\mathrm{erg \ cm^{-2} \ s^{-1}}]$  & $ 3.1 \times 10^{-12}$ &  $ 3.1 \times 10^{-12}$\\
(16) &  $L_{2-10} \ [\mathrm{erg \ s^{-1}}]$  & $ 2.9 \times 10^{41}$ & $ 2.6 \times 10^{41}$ \\
(17) &  $L_{10-50} \ [\mathrm{erg \ s^{-1}}]$  & $ 4.5 \times 10^{41} $ & $ 4.9 \times 10^{41}$ \\
(18) &  $ \mathrm{EW} \ [\mathrm{eV}]$ & $78$ & $62$ \\
  &  $\chi^2/$dof   & 61.9 / 78   &  62.1 / 78 
\enddata 
\tablenotetext{*}{
(1)~The line-of-sight hydrogen column density of the obscuring material.
(2)~The power-law photon index of the AGN transmitted component.
(3)~The power-law normalization of the AGN transmitted component at 1 keV.
(4)~The reflection strength.   
(5)~The scattering fraction.
(6)~The power-law photon index of the scattered component.
(7)~The temperature of the \textbf{apec} component.  
(8)~The normalization of the \textbf{apec} component.  
(9)~The flux time variability normalization of \Chandra/ACIS in 2013 July relative to FPMs.
(10)~The flux time variability normalization of \Chandra/ACIS in 2000 November relative to FPMs.
(11)~The flux time variability normalization of \XMM/EPIC-pn and MOSs in 2000 January relative to FPMs.
(12)~The instrumental cross normalization of EPIC-MOSs relative to FPMs.  
(13)~The instrumental cross normalization of EPIC-pn relative to FPMs.  
(14)~The observed flux in the 2--10 keV band.  
(15)~The observed flux in the 10--50 keV band.  
(16)~The de-absorbed AGN luminosity in the 2--10 keV band.  
(17)~The de-absorbed AGN luminosity in the 10--50 keV band.  
(18)~The equivalent width of the iron-K emission line with respect to the total continuum. \\} 
\end{deluxetable}

\subsection{NGC 833}

\subsubsection{Analytical Model}

\citet{tur01_hcg16} and \citet{sul14}, who analyzed the \XMM \ and
\Chandra \ spectra, respectively, found that NGC~833 has at least three
components: an absorbed cutoff power law, an unabsorbed power law, and
soft thermal components. We thus start from this model, which 
can reproduce our spectra including the \NuSTAR\ one with $\chi^2$/dof =
65.8/79. As a more realistic model, we also include a reflection
component accompanied by an iron-K emission line, which is known to
be commonly present in obscured AGNs (e.g., \citealt{nan94}; \citealt{kaw16a}).
The fit is 
improved ($\chi^2$/dof = 61.9/78) at the 95\% confidence level with 
an F-test. The adopted model
in the XSPEC terminology is expressed as:
\begin{eqnarray}
&& \mathbf{const0} * \mathbf{phabs}     \nonumber     \\  
&*& ( \mathbf{const1} * \mathbf{zphabs} *\mathbf{cabs}* \mathbf{zpowerlw} *\mathbf{zhighect}   \nonumber \\ 
&+& \mathbf{const1} * \mathbf{pexmon}  \nonumber \\
&+& \mathbf{const2} * \mathbf{zpowerlw} * \mathbf{zhighect}  \nonumber \\
&+&  \mathbf{apec}). 
\end{eqnarray}
The model consists of four components. The first term represents the
transmitted emission from the AGN, which is described by an
absorbed power law with a high-energy cutoff (primary component). We fix
a cutoff energy at 360 keV in \textbf{zhighect}, which cannot be
constrained from the data, for consistency with the torus model by
\citet{ike09} (see Section 3.1.2). The \textbf{cabs} model is
multiplied to take into account Compton scattering. The constant factor
(\textbf{const1}) is introduced to consider possible time variability of
the primary component among multiple observation periods; we set it
unity for the FPM spectrum, which is adopted as the flux reference.

The second term accounts for a reflection component from cold matter,
most likely the ``torus'' of the AGN. We utilize the \textbf{pexmon}
code (\citealt{nan07}), which calculates a Compton-reflection continuum from a
plane-parallel, semi-infinite cold matter (\textbf{pexrav},
\citealt{mag95}) including fluorescence lines.  The reflection strength
is defined as $R \equiv \Omega / 2 \pi$ ($\Omega$ is the solid angle of
the reflector), and we allow it to vary within a range of $-2 \le R < 0$
(a negative value means that the direct component is not included in the
model).  The inclination angle is fixed at 60$\degree$ as a representative
value, whereas the photon index and normalization are linked to those of
the primary component. We do not apply the same absorption as for the
primary emission (\textbf{zphabs}) to the reflection component, since it
does not improve the fit at $>$90\% confidence level.
The same time variability constant
\textbf{const1} as for the primary component is multiplied by the
reflection component, assuming that they roughly follow the long time
variability of the transmitted emission. 
If we instead assume that the reflection component is mainly produced by
matter located far from the black hole ($>$several pc) and is constant
over the whole observation epochs (i.e., \textbf{const1}$=$1), the
results do not change over the statistical uncertainties\footnote{This
is also true for NGC 835 (Section 3.2)}.

The third term mainly represents a scattered component of the AGN
emission, which is modeled with an unabsorbed cutoff power law.
Considering possible contribution from high mass X-ray binaries, we set
the photon index independent of that of the primary component. For
convenience, we tie the power law normalizations of the first and third
terms, and multiply a scattering fraction (\textbf{const2},
$f_\mathrm{scat}$) to the latter, which is defined as the ratio of the
unabsorbed fluxes at 1 keV between the primary and scattered
components. Since the scattering region is believed to be large
($\sim$kpc), we assume no time variability of this component among our
observation epochs. The fourth term describes optically-thin thermal
emission from the host galaxy. We also assume that this did not vary
among the observations because it likely originates from the starburst activity.

\begin{figure}[t!]
\begin{center}
\plotone{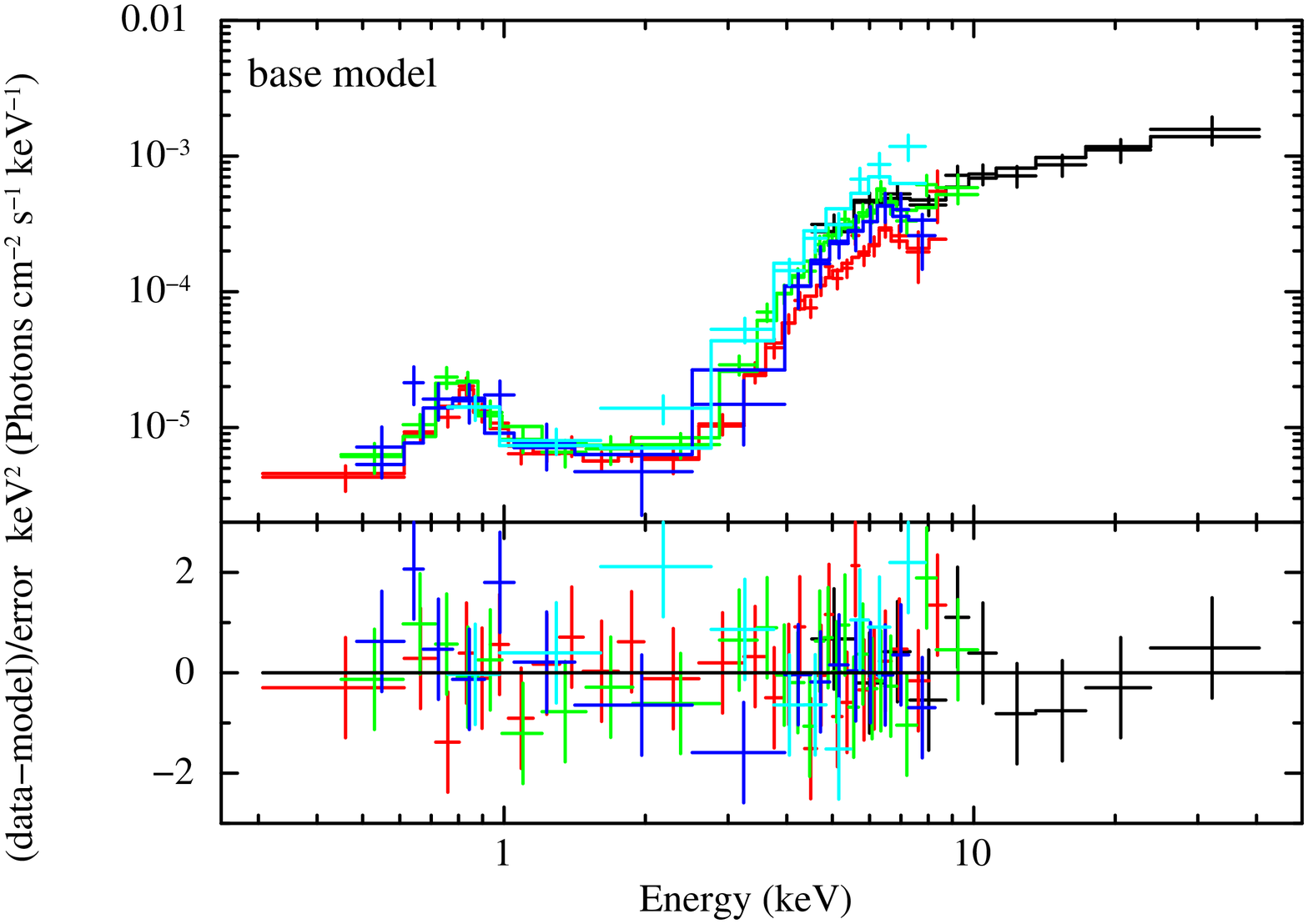}
\plotone{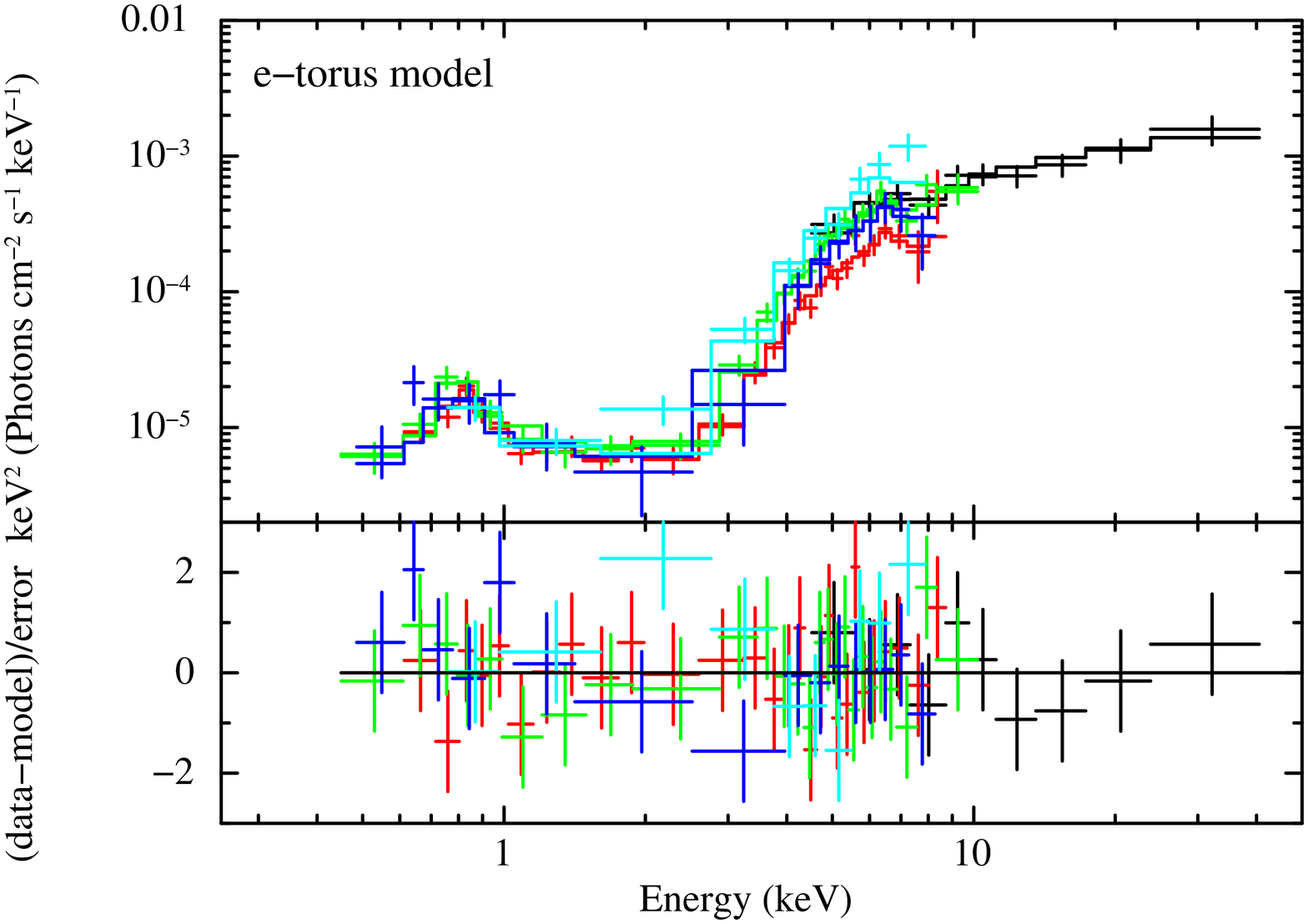}
\end{center}
\caption{Unfolded spectra of NGC~833 in units of $E I_E$ ($I_E$ is
the energy flux at the energy $E$).
The black, red, cyan, blue, and green crosses correspond to the
data of FPMs, ACIS (2013), ACIS (2000), EPIC-pn, and EPIC-MOSs,
respectively. The solid lines represent the best-fit model.}
\label{spec1}
\end{figure}

\begin{figure}[t!]
\begin{center}
\plotone{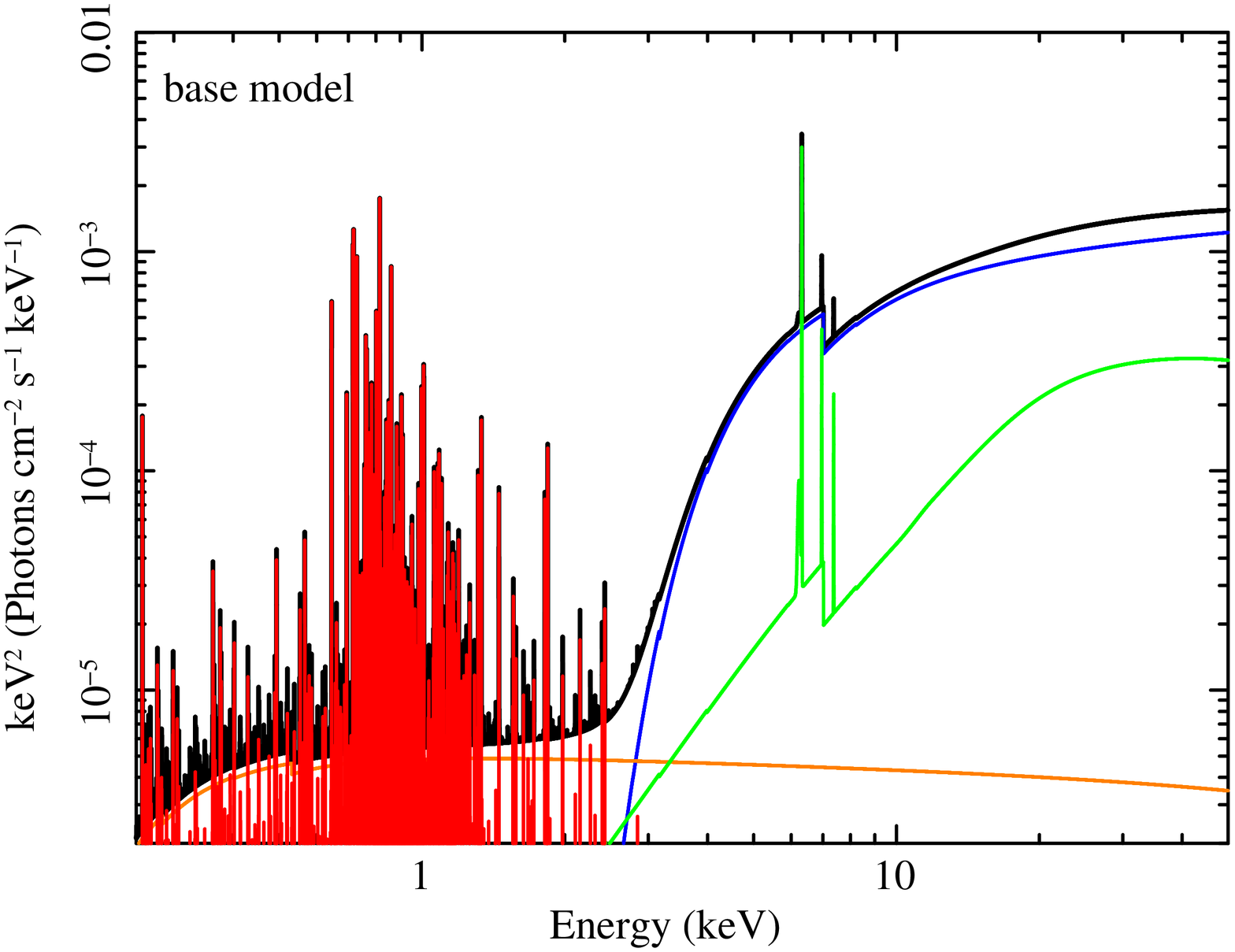}
\plotone{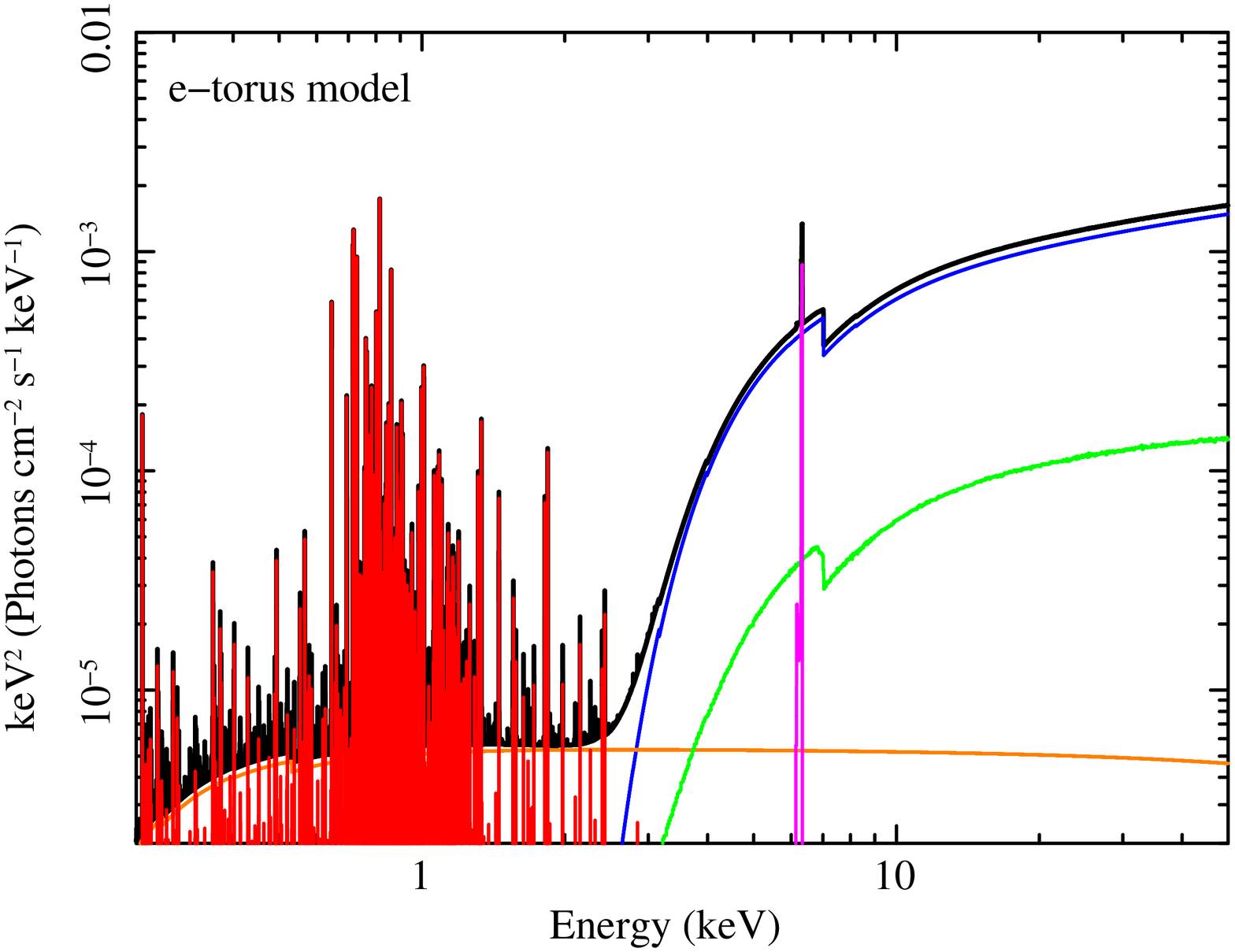}
\end{center}
\caption{Best-fit model of NGC~833 in units of $E I_E$.
The black, blue, green, magenta, orange, and red
lines represent the total, transmitted component, reflection component,
iron-K$\alpha$ emission line,
scattered component,
and optically-thin thermal component, respectively.
For the base model, the iron-K$\alpha$ line is merged
to the reflection component.}
\label{eem1}
\end{figure}

This model gives an acceptable fit of the combined spectra covering the
0.3--40 keV band ($\chi^2$/dof = 61.9/78).  Table~\ref{par1}
lists the best-fit parameters, the observed fluxes and de-absorbed
luminosities in the 2--10~keV and the 10--50 keV bands (based on the FPM
spectrum), and the equivalent width of the iron-K$\alpha$ emission line
with respect to the total continuum\footnote{To estimate the error in the
iron K$\alpha$ equivalent width, we fit the spectra by replacing the
\textbf{pexmon} component with a \textbf{pexrav} continuum plus a
narrow gaussian fixed at 6.4 keV.}.
The unfolded spectra and best-fit
model are plotted in Figures~\ref{spec1} and Figure~\ref{eem1},
respectively. We obtain the line-of-sight hydrogen column density of the
transmitted component of $N_\mathrm{H}^\mathrm{LS} =
2.68^{+0.37}_{-0.32} \times 10^{23}$ cm$^{-2}$.

\subsubsection{Torus Model}

To consider a realistic geometry of the torus, we fit the spectra with a
Monte-Carlo based numerical spectral model from a smooth torus developed
by \citet{ike09} (hereafter ``e-torus'' model).  The e-torus model
reproduces reflected spectra from a torus composed of constant-density
cold matter that has two cone-shaped holes along the polar axis (see
Figure~2 in \citealt{ike09}). The torus parameters are the ratio of the
inner ($r_\mathrm{in}$) and outer ($r_\mathrm{out}$) radii, which is
fixed at $r_\mathrm{in}/r_\mathrm{out} = 0.01$, the hydrogen column
density along the equatorial plane ($N_\mathrm{H}^\mathrm{Eq}$), and the
half-opening angle ($\theta_\mathrm{op}$ within a range of 0$\degree$--70$\degree$).
The primary X-ray spectrum is modeled by a power law
with a high energy cutoff at 360 keV. The other parameters are the photon
index ($\Gamma$ within a range of 1.5--2.5) and the inclination angle
($\theta_\mathrm{incl}$ within a range of 1$\degree$--89$\degree$). In this work,
we fix $\theta_\mathrm{incl}$ at 80$\degree$ to ensure a type-2 AGN based
on the optical classification, and $\theta_\mathrm{op}$ at 60$\degree$
as a representative value, both cannot be well constrained from our
data.
In the XSPEC terminology, the full spectral model utilizing the e-torus model
is expressed as it follows:
\begin{eqnarray}
&& \mathbf{const0} * \mathbf{phabs}     \nonumber     \\  
&*& ( \mathbf{const1} * \mathbf{torusabs} * \mathbf{zpowerlw} *\mathbf{zhighect}   \nonumber \\ 
&+& \mathbf{const1} * \mathbf{zpowerlw} * \mathbf{zhighect} \nonumber \\
&  & * \mathbf{mtable\{e\mathchar`-torus\_20161121\_2500M.fits\}}  \nonumber\\
&+& \mathbf{const1} * \mathbf{atable\{refl\_fe\_torus.fits\}}  \nonumber \\
&+& \mathbf{const2} * \mathbf{zpowerlw} * \mathbf{zhighect}  \nonumber \\
&+&  \mathbf{apec}). 
\end{eqnarray}
The model consists of five components: the transmitted component from
the AGN, the torus reflection component, the fluorescent iron-K$\alpha$
emission line, the scattered component multiplied by a scattering
fraction of $f_\mathrm{scat}$, and the optically-thin thermal component.
For a given torus geometry, we can convert the equatorial hydrogen
column density ($N_\mathrm{H}^\mathrm{Eq}$) into the line-of-sight one
($N_\mathrm{H}^\mathrm{LS}$) using Equation (3) in \citet{ike09},
which is represented by the \textbf{torusabs} model; for
$\theta_\mathrm{incl}=80\degree$ and $\theta_\mathrm{op}=60\degree$,
$N_\mathrm{H}^\mathrm{LS} \simeq N_\mathrm{H}^\mathrm{Eq}$.
The photon
indices are linked between the transmitted and reflection
components. The power-law normalizations are linked among the
transmitted, reflection, and scattered components.
As is the case with the analytical model, the cross calibration
(\textbf{const0}) and time variability (\textbf{const1}) are defined
with respect to the \NuSTAR \ spectrum. We do not use the energy band 
below 0.45 keV, where the e-torus model is unavailable.

We find that this model also well reproduces the combined spectra of
NGC~833 covering the 0.45--40 keV band ($\chi^2$/dof =
62.1/78). Table~\ref{par1} lists the best-fit parameters, the observed
fluxes and intrinsic luminosities in the 2--10~keV and the 10--50 keV
energy band (based on the FPM spectrum), and the equivalent width of the
iron-K$\alpha$ emission line.  Figure~\ref{spec1} and Figure~\ref{eem1}
represent the unfolded spectra and best-fit model, respectively. The
line-of-sight column density is found to be $N_\mathrm{H}^\mathrm{LS} =
2.54^{+0.31}_{-0.25} \times 10^{23}$ cm$^{-2}$, which is consistent with
the result from the analytical model.

\subsection{NGC 835}

\subsubsection{Analytical Model}

We adopt essentially the same model as for NGC~833. It is expressed as: 
\begin{eqnarray}
&& \mathbf{const0} * \mathbf{phabs}     \nonumber     \\  
&*& ( \mathbf{const1} * \mathbf{zphabs} *\mathbf{cabs}* \mathbf{zpowerlw} *\mathbf{zhighect}   \nonumber \\ 
&+& \mathbf{const1} * \mathbf{pexmon}  \nonumber \\
&+& \mathbf{const2} * \mathbf{zpowerlw} * \mathbf{zhighect}  \nonumber \\
&+&  \mathbf{apec1} + \mathbf{apec2}). 
\end{eqnarray}
A difference from NGC~833 is that we include two optically thin
thermal emission components with different temperatures, which are
required from the data.  They are likely to originate from the core and
outer-ring star-forming regions as discussed in \citet{tur01_hcg16}.

As mentioned in Section~1, \citet{sul14} and \citet{gon16} suggested that the
X-ray spectrum of NGC~835 was variable on timescales of months to years
due to changes in the luminosity or absorption. 
Accordingly, we first
allow the column density $N_\mathrm{H}^\mathrm{LS}$ to vary among the
\NuSTAR \ (2015 September), \Chandra \ (2013 July), \Chandra \ (2000 November),
and \XMM \ (2000 January) observations. We find that the column densities obtained are
consistent between the \NuSTAR\ and \Chandra\ (2013) data, whereas that
measured with \XMM\ ($N_\mathrm{H2}^\mathrm{LS}$) is significantly
larger than the one measured with \Chandra\ (2013) at a 90\%
confidence level. Unfortunately, the column density in 2000 November
cannot be well constrained owing to 
the limited photon statistics of the \Chandra \ data.
Thus, in the final model, we tie the column densities of the \NuSTAR\ and
\Chandra \ (2013) spectra together
($N_\mathrm{H1}^\mathrm{LS}$), and also those of the \XMM \ and \Chandra \
(2000) spectra ($N_\mathrm{H2}^\mathrm{LS}$).
Even if we instead link the column density of the \Chandra\ (2000)
spectrum to those of the \NuSTAR\ and \Chandra\ (2013) spectra, the
column density inferred from the \XMM \  spectrum does not change significantly and our 
conclusions are not affected.

This model well reproduces all the spectra covering the 0.3--49 keV band 
($\chi^2$/dof = 150.6/162).
We confirm that inclusion of the reflection component (the second term of Equation (3))
significantly improves the fit ($\chi^2$/dof = 164.7/163 without it).
The fitting results are summarized in 
Table~\ref{par2}. 
Figures~\ref{spec2} and \ref{eem2} plot the unfolded spectra and the
best-fit model, respectively.
We find that the column density
$N_\mathrm{H}^\mathrm{LS}$ varied from 
$5.2^{+1.9}_{-1.3} \times 10^{23}$ cm$^{-2}$ (2000) to 
$2.9^{+0.3}_{-0.2} \times 10^{23}$ cm$^{-2}$ (2013/2015).  
In addition, the 2--10~keV intrinsic luminosity ($L_\mathrm{X}$) was
also variable: it was $\approx 1.1 \times 10^{41}$ erg s$^{-1}$ in 2000 January
(\XMM), $\approx 2.5 \times 10^{41}$ erg s$^{-1}$ in
2000 November (\Chandra), $\approx 5.8 \times 10^{41}$ erg s$^{-1}$ in
2013 (\Chandra), and $\approx 3.4 \times 10^{41}$ erg s$^{-1}$ in 2015
(\NuSTAR).

\begin{deluxetable}{cccc}
\tablecaption{Best-fit Parameters of NGC~835 \label{par2}}
\tablewidth{0.5\textwidth}
\tablehead{
 Note$^{*}$ & Parameter & base model & e-torus model
}
\startdata
(1) &  $ N_\mathrm{H1}^\mathrm{LS} \  [10^{23} \ \mathrm{cm^{-2} }]$   & $2.94^{+0.27}_{-0.24}$ & $2.86^{+0.16}_{-0.15}$ \\
(2) &  $\Gamma_\mathrm{AGN}$ &  $1.50 \pm 0.17$  & $1.50\ (<1.57)$ \\
(3) & $A_\mathrm{AGN} \ [10^{-4} \ \mathrm{keV^{-1}\ cm^{-2} \ s^{-1}}]$   & $4.5^{+2.5}_{-1.6}$ &  $4.36^{+0.86}_{-0.33}$ \\
(4) &  $|R|$ & $0.27^{+0.14}_{-0.13}$ & -- \\
(5) & $f_\mathrm{scat}\ [\%] $   &  $2.1^{+1.3}_{-0.8}$  &  $2.19^{+0.46}_{-0.50}$  \\
(6) & $\Gamma_\mathrm{scat}$ & $2.17^{+0.18}_{-0.21}$ & $2.00^{+0.26}_{-0.31}$\\
(7) & $k \mathrm{T}_1 \ [\mathrm{keV}] $   &  $0.40^{+0.07}_{-0.05}$  & $0.39^{+0.06}_{-0.04}$ \\  
(8) &  $ A_\mathrm{apec1}\ [10^{-5} \ \mathrm{cm^{-5}}]$  &  $0.98^{+0.19}_{-0.21}$  &  $1.04^{+0.22}_{-0.24}$ \\
(9) &  $k \mathrm{T}_2 \ [\mathrm{keV}] $   &  $0.89^{+0.06}_{-0.08}$  &  $0.89^{+0.07}_{-0.08}$ \\  
(10) &  $ A_\mathrm{apec2}\ [10^{-5} \ \mathrm{cm^{-5}}]$  &  $0.77 \pm 0.12$  &  $0.78^{+0.13}_{-0.12}$ \\
(11) &  $N_\mathrm{Ch13}^\mathrm{time}$  & $1.72^{+0.23}_{-0.19}$ & $1.65^{+0.16}_{-0.14}$ \\
(12) &  $N_\mathrm{Ch00}^\mathrm{time}$  & $0.73^{+0.79}_{-0.39}$ & $0.68^{+0.66}_{-0.34}$ \\
(13) & $N_\mathrm{XMM}^\mathrm{time}$  & $0.33^{+0.19}_{-0.11}$ & $0.30^{+0.15}_{-0.08}$\\
(14) &  $N_\mathrm{MOS}$ & $1.25 \pm 0.08$ & $1.23 \pm 0.08$\\
(15) &  $N_\mathrm{pn}$  & $ 1.11^{+0.07}_{-0.06}$  & $1.09^{+0.07}_{-0.06}$ \\
(16) &  $N_\mathrm{H2}^\mathrm{LS} \  [10^{23} \ \mathrm{cm^{-2} }]$  & $5.2^{+1.9}_{-1.3}$ & $5.0^{+1.5}_{-1.0}$\\
(17) &  $F_{2-10} \ [\mathrm{erg \ cm^{-2} \ s^{-1}}]$  & $ 7.2 \times 10^{-13}$ & $ 7.5 \times 10^{-13}$ \\
(18) &  $F_{10-50} \ [\mathrm{erg \ cm^{-2} \ s^{-1}}]$  & $  4.7 \times 10^{-12}$ &  $ 4.3 \times 10^{-12}$ \\
(19) &  $L_{2-10} \ [\mathrm{erg \ s^{-1}}]$  & $ 3.4 \times 10^{41} $ &  $ 3.3 \times 10^{41}$ \\
(20) &  $L_{10-50} \ [\mathrm{erg \ s^{-1}}]$  & $ 7.0 \times 10^{41}$ & $ 6.9 \times 10^{41}$  \\
(21) &  $ \mathrm{EW} \ [\mathrm{eV}]$ & 80 & 69 \\
  &  $\chi^2/$dof   &  150.6 / 162   &  152.0 / 156
\enddata 
\tablenotetext{*}{
(1)~The line-of-sight hydrogen column density of the obscuring material in 
2013 (\Chandra) and 2015 (\NuSTAR).
(2)~The power-law photon index of the AGN transmitted component.
(3)~The power-law normalization of the AGN transmitted component at 1 keV.
(4)~The reflection strength.
(5)~The scattering fraction.
(6)~The power-law photon index of the scattered component.  
(7)(9)~The temperatures of the \textbf{apec} components.  
(8)(10)~The normalizations of the \textbf{apec} components. 
(11)~The flux time variability normalization of \Chandra/ACIS in 2013 July relative to FPMs.
(12)~The flux time variability normalization of \Chandra/ACIS in 2000 November relative to FPMs.
(13)~The flux time variability normalization of \XMM/EPIC-pn and MOSs in 2000 January relative to FPMs.
(14)~The instrumental cross normalization of EPIC-MOS relative to FPMs.
(15)~The instrumental cross normalization of EPIC-pn relative to EPIC-MOSs.  
(16)~The line-of-sight hydrogen column density of the torus in 2000
 (\XMM ).
(17)~The observed flux in the 2--10 keV band.  
(18)~The observed flux in the 10--50 keV band.  
(19)~The de-absorbed AGN luminosity in the 2--10 keV band.  
(20)~The de-absorbed AGN luminosity in the 10--50 keV band.  
(21)~The equivalent width of the iron-K emission line with respect to the total continuum. } 
\end{deluxetable}

\subsubsection{Torus Model}

We also adopt essentially the same torus model as for
NGC~833
expressed as: 
\begin{eqnarray}
&& \mathbf{const0} * \mathbf{phabs}     \nonumber     \\  
&*& ( \mathbf{const1} * \mathbf{torusabs} * \mathbf{zpowerlw} *\mathbf{zhighect}   \nonumber \\ 
&+& \mathbf{const1} * \mathbf{zpowerlw} * \mathbf{zhighect} \nonumber \\
&  & * \mathbf{mtable\{e\mathchar`-torus\_20161121\_2500M.fits\}}  \nonumber\\
&+& \mathbf{const1} * \mathbf{atable\{refl\_fe\_torus.fits\}}  \nonumber \\
&+& \mathbf{const2} * \mathbf{zpowerlw} * \mathbf{zhighect}  \nonumber \\
&+&  \mathbf{apec1} + \mathbf{apec2}). 
\end{eqnarray}
We include two thermal components as in the case of the analytical model.
The inclination angle and half-opening angle are fixed at 80$\degree$ and
60$\degree$, respectively, because they cannot be constrained from the
data. To take into account the variability in the line-of-sight column
density, we simply allow the $N_\mathrm{H}^\mathrm{Eq}$ parameters
to be independent between the 
observations in 2000 and 2013/2015.  
This would be an unrealistic assumption if the overall structure of the
torus was stable on a timescale of years and only the local
absorption was variable e.g., due to a passage of a clump across 
the line-of-sight.
Nevertheless, even if we make only the line-of-sight absorption
variable among the two epochs (2000 and 2013/2015), 
being decoupled 
from a constant $N_\mathrm{H}^\mathrm{Eq}$ value,
the results do not change over the statistical errors.

\begin{figure}[t!]
\begin{center}
\plotone{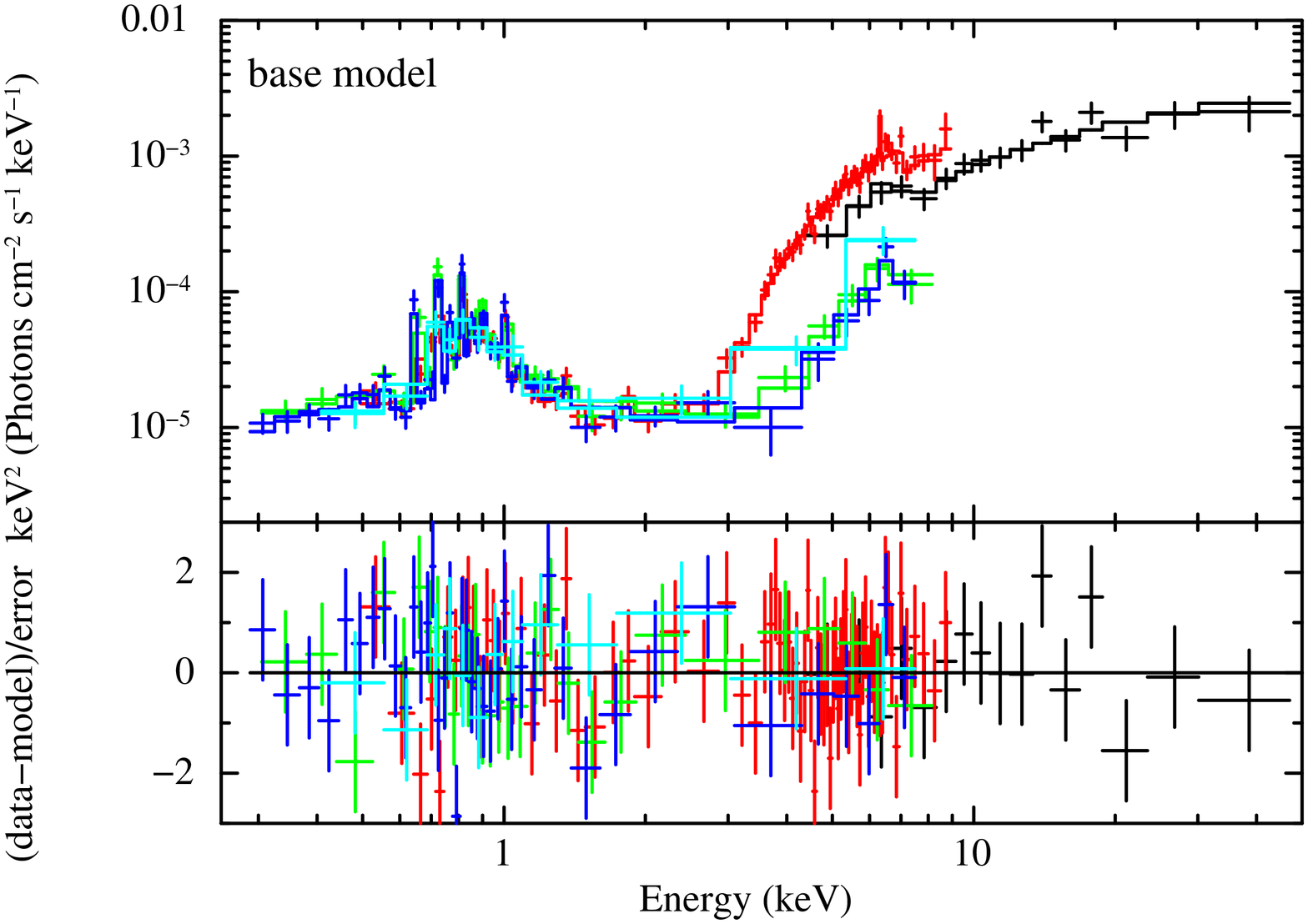}
\plotone{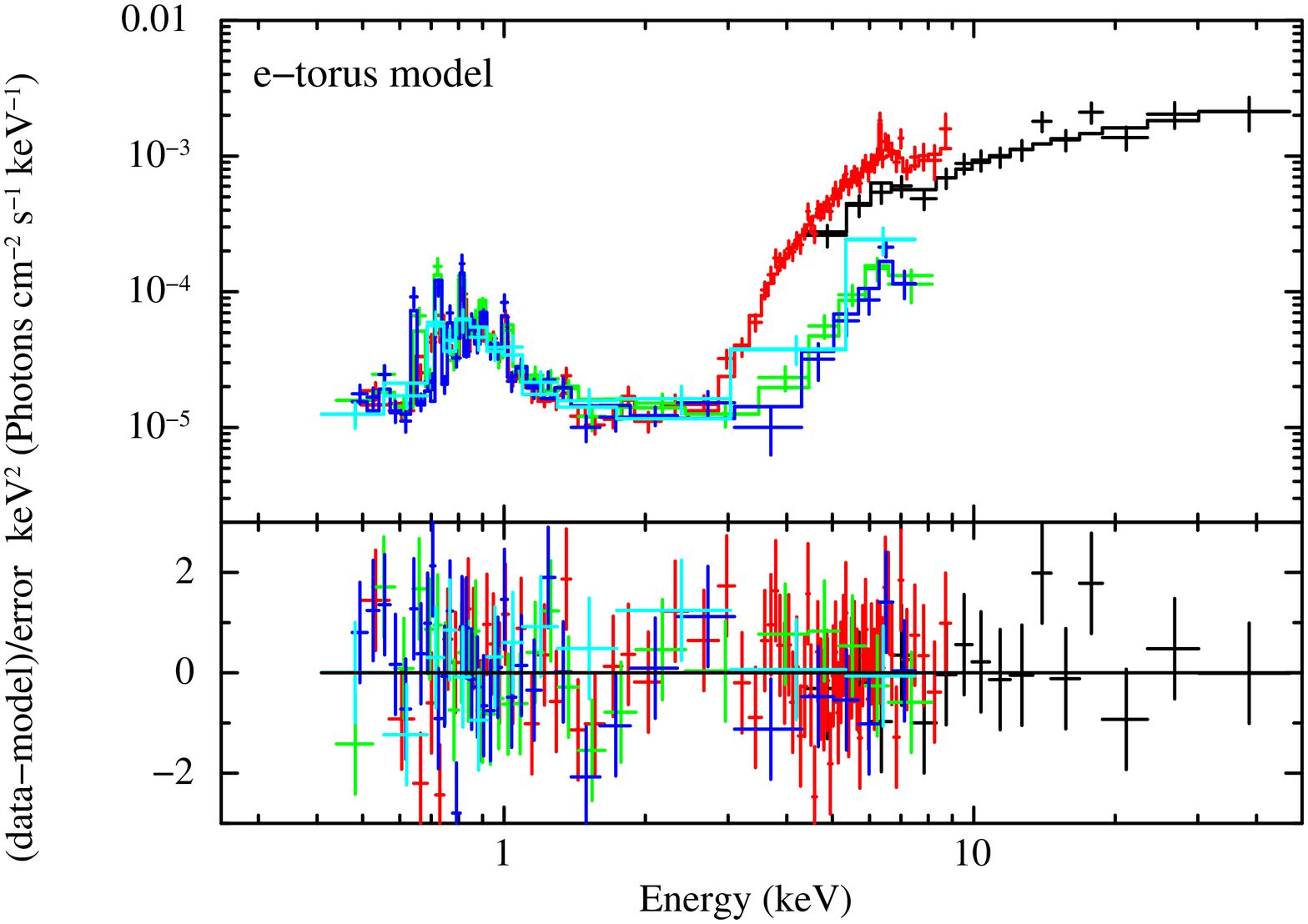}
\end{center}
\caption{Unfolded spectra of NGC~835 in units of $E I_E$.
The black, red, cyan, blue, and green crosses are the
data of FPMs, ACIS (2013), ACIS (2000), EPIC-pn, and EPIC-MOSs,
respectively. The solid lines represent the best-fit model.}
\label{spec2}
\end{figure}

\begin{figure}[t!]
\begin{center}
\plotone{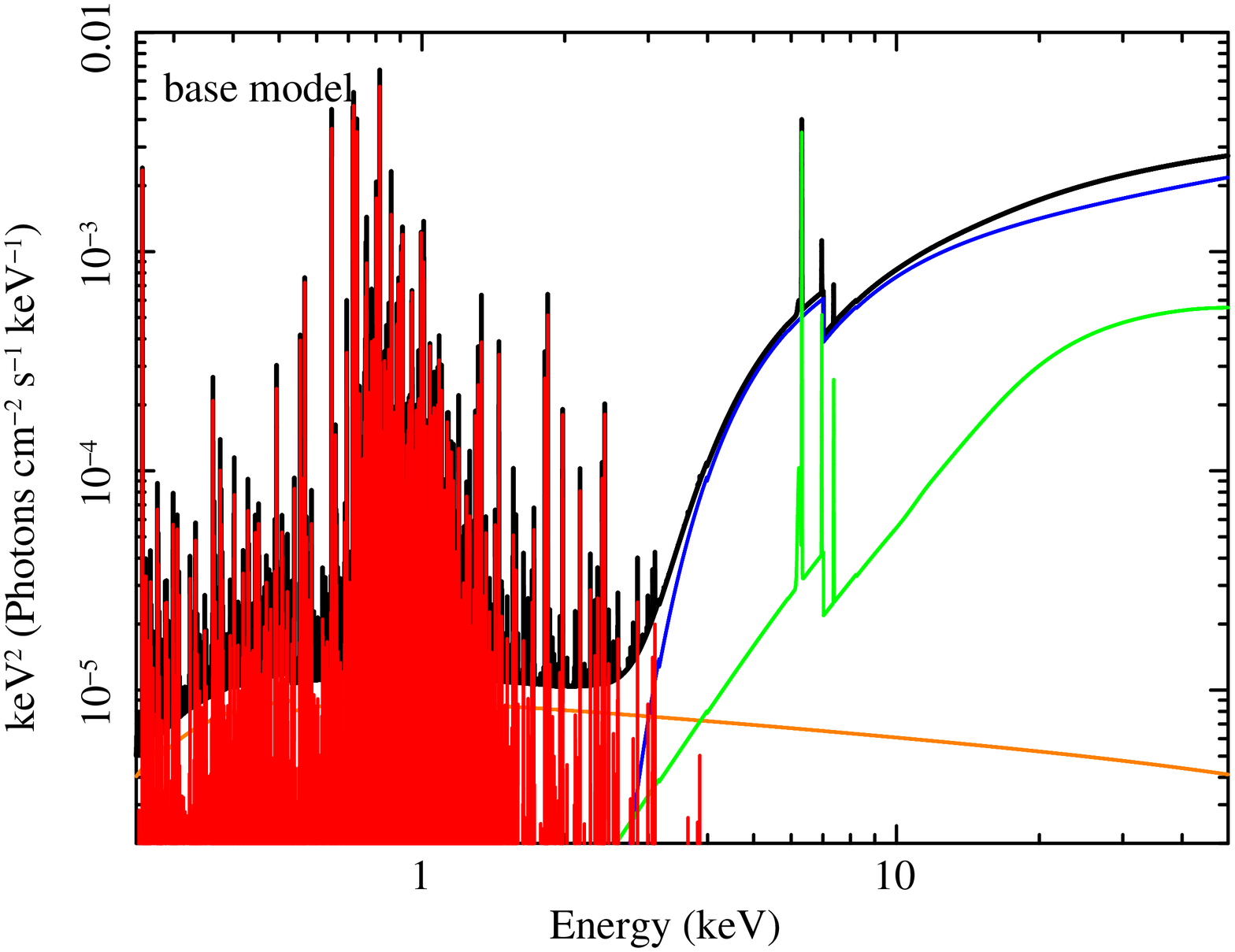}
\plotone{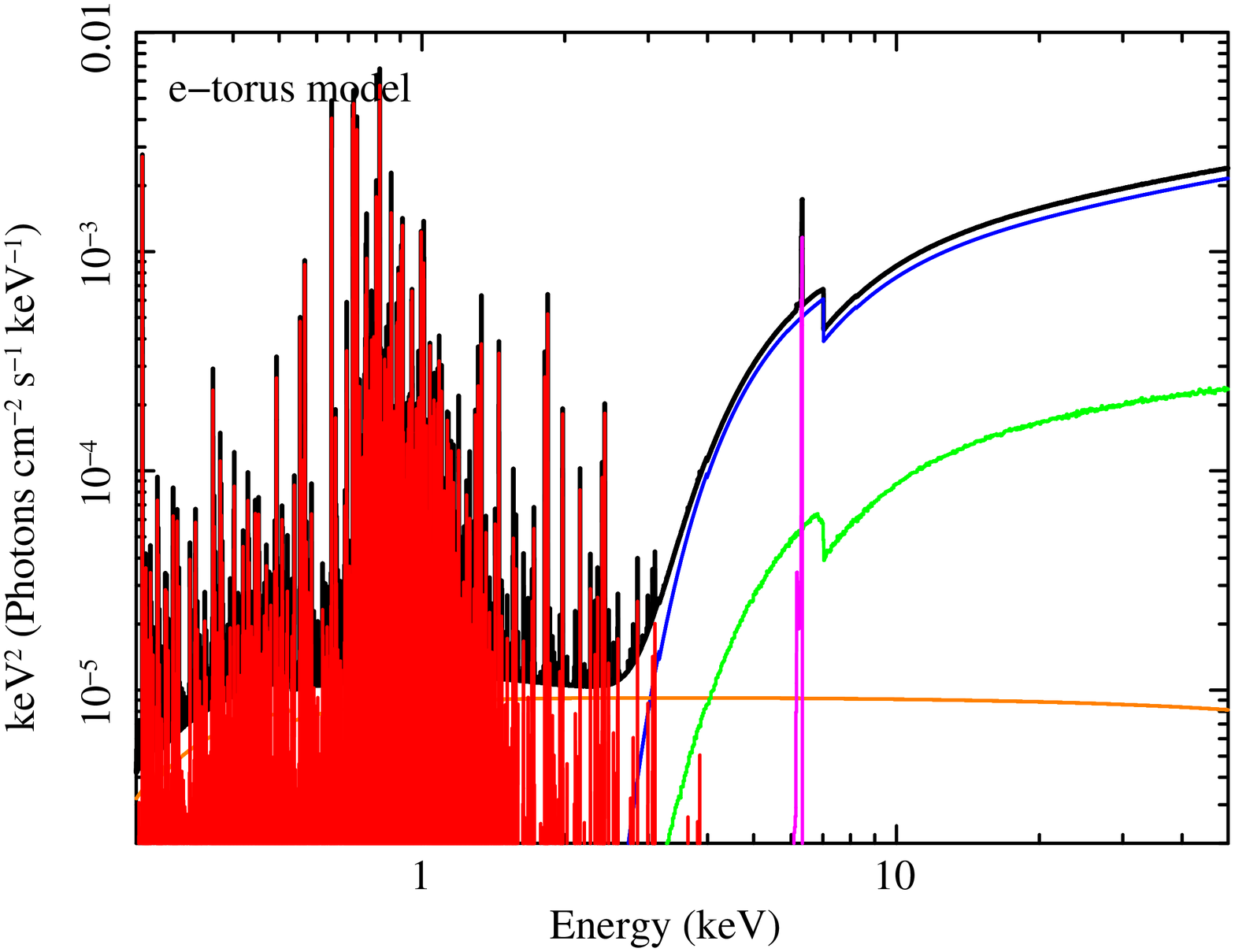}
\end{center}
\caption{Best-fit model of NGC~835 in units of $E I_E$.
The black, blue, green, magenta, orange, and red
lines represent the total, transmitted component, reflection component,
iron-K$\alpha$ emission line,
scattered component,
and optically-thin thermal component, respectively.
For the base model, the iron-K$\alpha$ line is merged
to the reflection component.}
\label{eem2}
\end{figure}

The model also well reproduces all the spectra of NGC 835
covering the 0.45--49 keV band ($\chi^2$/dof = 152.0/156).
The results are summarized in Table~\ref{par2}, and 
the unfolded spectra and the best-fit model are plotted in Figures~\ref{spec2}
and \ref{eem2}, respectively.
It is found that the line-of-sight column density $N_\mathrm{H}^\mathrm{LS}$ changed
from $5.0^{+1.5}_{-1.0} \times 10^{23}$ cm$^{-2}$ (\XMM/\Chandra)
to $(2.9 \pm 0.2) \times 10^{23}$ cm$^{-2}$ (\NuSTAR /\Chandra). 
The intrinsic 2--10 keV luminosity was also variable: 
$\approx 1.0 \times 10^{41}$ erg s$^{-1}$ in 2000 January (\XMM),
$\approx 2.2 \times 10^{41}$ erg s$^{-1}$ in 2000 November (\Chandra),
$\approx 5.4 \times 10^{41}$ erg s$^{-1}$ in 2013 (\Chandra),
and $\approx 3.3 \times 10^{41}$ erg s$^{-1}$ in 2015 (\NuSTAR).
These results are consistent with those of the analytic model (see
Section~3.2.1).

\section{Hard X-ray Constraints on AGNs in NGC~838 and NGC~839}

As we mention in Section~2.1, NGC~838 and NGC~839 are not detected with \NuSTAR\
in the 3--24 keV band, with the upper flux limits consistent with the
extrapolation from the spectrum determined at energies below 10 keV.
This suggests that their AGNs, if present, must be very weak as
suggested by the previous works (\citealt{tur01_hcg16}; \citealt{sul14}), and/or are subject to
extremely heavy obscuration, e.g., $N_\mathrm{H} \gg 10^{25}$ cm$^{-2}$, from which even
hard X-rays cannot escape.

Here we evaluate the upper limits of the intrinsic luminosities of
possible AGNs that may be present in these two galaxies, using the
\NuSTAR\ data. According to \citet{sul14}, the \Chandra \ spectra of
both galaxies are well reproduced by a model consisting of an
optically-thin thermal plasma and emission from HMXBs expected from the
star forming activities. Subtracting the extrapolated fluxes of these
components, we finally obtain the 3$\sigma$ upper limits of 0.00050 and
0.00021 counts s$^{-1}$ in the \NuSTAR\ 8--24 keV band for NGC~838 and
NGC~839, respectively, for the AGN contribution. To convert these count
rates into the intrinsic luminosities, we utilize the same torus model
as described in Section~3.1.2 and 3.2.2 (without the scattered and thermal components). We
assume an inclination angle of 80$\degree$, a half-opening angle of 60$\degree$,
and a photon index of 1.9. Table~\ref{par3} lists the upper
limits of the intrinsic 2--10 keV luminosities of AGNs in NGC~838 and
NGC~839 for three assumed column densities. Even in an extreme case of
$N_\mathrm{H} = 10^{25}$ cm$^{-2}$, the AGN must be less luminous than a few times
$10^{42}$ erg s$^{-1}$ (2--10 keV).

Our results support the argument by \citet{sul14} that NGC~838 and NGC~839 are
starburst-dominant galaxies without luminous AGNs, which is consistent
with their optical classifications (see \citealt{vog13} for a detailed
study). The issue whether the hard X-ray component below 10 keV observed
from NGC~839 mainly originates from a very low-luminosity AGN (\citealt{tur01_hcg16}) or
from HMXBs (\citealt{sul14}) is, however, still left for an open question.

\tabletypesize{\footnotesize}
\begin{deluxetable}{ccc}[tb]
\tablecaption{Upper Limits of Intrinsic AGN luminosities of NGC~838 and
 NGC~839 obtained with \NuSTAR  \label{par3}}
\tablewidth{0.5\textwidth}
\tablehead{
    & \multicolumn{2}{c}{$L_{2-10}$ [erg s$^{-1}$]} \\
\cline{2-3}
$ \log N_\mathrm{H}^\mathrm{Eq}$ [cm$^{-2}$] & NGC 838 &  NGC 839
}
\startdata
24 & $< 5.3 \times 10^{40}$  & $< 3.0 \times 10^{40}$ \\
24.5 & $< 3.0 \times 10^{41}$  & $< 1.7 \times 10^{41}$ \\
25 & $< 3.1 \times 10^{42}$  & $< 1.8 \times 10^{42}$
\enddata
\tablecomments{
$N_\mathrm{H}^\mathrm{Eq}$ is an equatorial hydrogen column density of 
the torus.
$L_{2-10}$ denotes the 2--10 keV intrinsic AGN luminosity.
}
\end{deluxetable}

\section{Discussion}

We have analyzed the first hard X-ray ($>10$ keV) imaging data of the
compact group HCG~16 observed with \NuSTAR. NGC 833 and NGC 835 are
significantly detected with \NuSTAR, whereas we obtain the tightest flux
upper-limit for NGC 838 and NGC 839 above 10 keV.  Simultaneous broadband
(0.3--50 keV) spectral analysis utilizing the \NuSTAR, \Chandra, and
\XMM\ data, with a total exposure of 310.6 ks, has enabled us to best
constrain the X-ray properties of NGC~833 and NGC 835. In the analysis,
we have carefully corrected for contamination to the \NuSTAR\ spectrum
of one target from the other. In this section, we focus on the results
of NGC~833 and NGC~835 and discuss their implications,

It has been revealed that both galaxies contain moderately obscured
LLAGNs ($N_{\rm H} \approx 3\times10^{23}$~cm$^{-2}$ and $L_{2-10}
\approx 3\times10^{41}$ erg s$^{-1}$ based on the \NuSTAR\
spectra), confirming the previous reports by \citet{tur01_hcg16}
and \citet{sul14}. Their spectra
are well reproduced with both the analytical model and the numerical
torus model. We confirm that the two models give very similar results on
the photon index, line-of-sight absorption, and intrinsic luminosity for
each target. Hereafter we refer to the values obtained with the torus
model, where a realistic geometry is considered, unless otherwise stated.

\subsection{Torus Structure of LLAGNs in NGC 833 and NGC 835}

The fitting results with the analytical model show that the reflection
strength from cold matter is not strong in both targets, $R=0.31^{+0.31}_{-0.26}$ for
NGC 833 and $R=0.27^{+0.14}_{-0.13}$ for NGC 835. This suggests that  
their tori are only moderately developed. In fact, the spectra are also
consistent with the torus model with an opening angle fixed at 60$\degree$.
Here we compare this result with the previous ones obtained for isolated
(non-interacting) LLAGNs.

\citet{kaw16b} studied broadband X-ray spectra of nearby
LLAGNs observed with Suzaku. Using the luminosity
ratio of the iron-K$\alpha$ line to the hard X-ray (10--50 keV)
continuum as a good indicator of the torus covering fraction
(\citealt{ricci14}), they suggested that the Eddington ratio $\lambda_\mathrm{Edd}$
would be a key parameter that determines the torus structure of LLAGNs;
at $\lambda_\mathrm{Edd} > 2 \times 10^{-4}$ its solid angle is large, whereas
$\lambda_\mathrm{Edd} < 2 \times 10^{-4}$ it is significantly smaller.

The $L_{\rm K\alpha}/L_{10-50}$ ratios are found to be
$2.3^{+3.4}_{-1.4} \times 10^{-3}$ for NGC~833
and $2.0^{+0.6}_{-0.5} \times 10^{-3}$ for NGC~835.
The bolometric luminosities of NGC~833 and NGC~835
in 2015 are estimated to be 
$2.3 \times 10^{42}$ erg s$^{-1}$ and $3.0 \times 10^{42}$ erg s$^{-1}$, 
respectively, which are converted from the 2--10 keV
intrinsic luminosities by adopting to equation (21) of
\citet{mar04}. Using the SMBH mass versus galaxy velocity-dispersion
correlation of \citet{tre02}, we estimate the black hole mass of NGC~833
to be $M_\mathrm{BH} = 10^{8.0} M_\odot$ with the stellar velocity
dispersion available in the Hypercat database (\citealt{pat97},
http://www-obs.univ-lyon1.fr/hypercat).  This yields an Eddington ratio
of $\lambda_{\rm Edd}=1.6 \times 10^{-4}$ for NGC~833. The value of $L_{\rm
K\alpha}/L_{10-50}$ is consistent with what found by \citet{kaw16b}
for the range of $\lambda_\mathrm{Edd}$,
locating it around the transition region between
the well-developed tori at high $\lambda_\mathrm{Edd}$ and underdeveloped ones at
low $\lambda_\mathrm{Edd}$. 
Although the stellar velocity dispersion (hence SMBH mass) of
NGC~835 is unavailable, we infer that its Eddington ratio is similar to
that of NGC~833 on the basis of the $L_{\rm K\alpha}/L_{10-50}$ value.
A possibility is that the galaxy interaction has just
triggered new AGN activities in both objects.

\subsection{Origin of Flux Variability of NGC 835}

We have confirmed the flux variability of NGC 835 below 10 keV between
2000 (\XMM/\Chandra ) and 2013 (\Chandra ) found by \citet{sul14}. 
Our spectral fitting results including the \NuSTAR\ data in 2015 suggest
that a significant variation in the line-of-sight absorption of $\Delta
N_{\rm H}\approx 2\times10^{23}$ cm$^{-2}$ is required to explain the
spectral difference between 2000 (observed with \XMM \ and \Chandra ) and 2013/2015
(\Chandra/\NuSTAR), in addition to changes in the intrinsic luminosity
by a maximum factor of $\sim$5 among the three epochs (Table~6).
Such large long-term variability in the intrinsic luminosity is often
observed in nearby LLAGNs (e.g., \citealt{kaw16b}).

The variable absorption on a timescale of years may be associated with a
transit of high density clouds in the clumpy torus, as already discussed
by \citet{sul14} and \citet{gon16}. \citet{mark14} detected such events with column
densities of $\Delta N_\mathrm{H} \sim 10^{22-23}$~cm$^{-2}$ from
$\sim$10 nearby Seyferts by analyzing their long-term monitoring X-ray
light curves. They estimated the distance of clumps from the SMBHs to be
$0.3-140 \times 10^4 R_\mathrm{g}$ ($R_\mathrm{g} \equiv \mathrm{G
M}_\mathrm{BH}\mathrm{c}^{-2}$ is the gravitational radius).  Because of
the limited number of multiple observations, we are not able to measure
the clump crossing-time in NGC~835 except its lower limit, $\sim$0.7 day
(duration of the \XMM\ observation in 2000
January; we conservatively adopt this value because the column
density of the \Chandra\ spectrum in 2000 January is highly uncertain).
With this lower limit, the clump distance is constrained to be $>1.0
\times 10^3 R_\mathrm{g}$ by assuming the SMBH mass of $\log
M_\mathrm{BH}/M_\odot = 8$, the ionizing luminosity of $L_\mathrm{ion}=
2.5 \times 10^{41}$ erg s$^{-1}$, and the ionization parameter of $\log
\xi = 0$, according to equation (3) in \citet{lam03}.
This lower limit corresponds to a typical location of broad line regions
but does not contradict the results by \citet{mark14}.

\begin{figure}[tb]
\begin{center}
\plotone{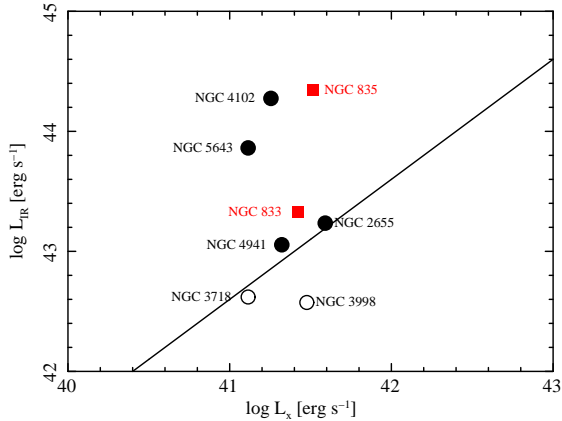}
\end{center}
\caption{Correlation between the 2--10 keV AGN luminosity and the
infrared (8--1000$\mu$m) luminosity of local LLAGNs. Our targets are
marked with red-filled squares, and LLAGNs in \citet{kaw16b} are
marked with black-filled circles for higher $\lambda_\mathrm{Edd}$
and black-open circles for lower $\lambda_\mathrm{Edd}$.
The solid line represents the average relation for PG QSOs (\citealt{ten10}).}
\label{x_ir}
\end{figure}

\subsection{Infrared to X-ray Luminosity Relation of Local LLAGNs}

In this subsection we briefly investigate possible connection between
the AGN and star-forming activities in local galaxies hosting
LLAGNs. Figure~\ref{x_ir} plots the relation between AGN X-ray
luminosity in the 2--10 keV band ($L_\mathrm{X}$) and the infrared luminosity in
the 8--1000 $\mu$m band ($L_\mathrm{IR}$) for NGC~833 and NGC~835 (red-filled
squares). For comparison, we also plot the data of non-interacting
LLAGNs, NGC~2655, NGC~3718, NGC~3998, NGC~4102, NGC~4941, and NGC~5643
(black-filled circles and black-open circles) reported in \citet{kaw16b}, for which the
\textit{Infrared Astronomical Satellite} (\textit{IRAS}) photometric data
are available\footnote{ The 8--1000 $\mu$m luminosities are calculated
as $L_\mathrm{IR} \ [\mathrm{erg \ s}^{-1}]= 2.1 \times 10^{39} \times
D^2 \times (13.48 \times f_{12} +5.16 \times f_{25} + 2.58 \times f_{60}
+ f_{100})$ (\citealt{san96}), where $D$ is a luminosity distance in
unit of Mpc, and $f_{12}$, $f_{25}$, $f_{60}$, and $f_{100}$ are the
\textit{IRAS} fluxes at 12, 25, 60, and 100 $\mu$m in unit of Jy,
respectively.}. The average relation obtained for PG QSOs by
\citet{ten10} is represented by the solid line as reference.

The X-ray and infrared luminosities are indicators of the mass accretion
rate onto the SMBH and star forming rate in the host galaxy (after subtracting AGN
contribution), respectively. Assuming a bolometric correction factor of
$L_{\rm bol}/L_\mathrm{2-10}\sim10$ \citep{mar04}, we find that the ratio of
the AGN bolometric luminosity to the observed infrared luminosity is
much smaller than unity in these objects except for NGC~3998.  Since the
torus in NGC~3998 is known to be almost absent (e.g., \citealt{kaw16b}), the
hot dust emission from the torus in the infrared band is expected to be
very small compared to the AGN bolometric luminosity.  Hence, we can
regard that AGN contribution to the infrared luminosities is negligible
in all the objects plotted here.

Figure~\ref{x_ir} indicates large diversity in the $L_\mathrm{X}$ versus
$L_\mathrm{IR}$ relation (hence, mass accretion rate versus star forming
rate relation) in the local LLAGNs, regardless if the galaxy is in
interacting systems (red filled squares) or not (black filled circles
and black open circles). All the objects are spiral galaxies and the
star forming rate is dominated by that in the disk. Our results are not
surprising, given the fact that there is little correlation between SMBH
mass and stellar mass in the {\it disk} component in the local universe, 
unlike the tight correlation between SMBH mass and bulge mass 
(see \citealt{kor13}). In fact, we find no evidence that the torus
structure is largely different between NGC 833 and NGC 835 despite of
the large difference in the $L_\mathrm{IR}$/$L_\mathrm{X}$ ratio,
supporting the general idea that the star forming activity in the disk
is not strongly coupled to the mass accretion process in the nucleus.

\section{Conclusion}

In this paper, 
we report the first hard X-ray ($>$10 keV) observation 
of HCG~16 performed with \NuSTAR. We particularly study
the broadband X-ray spectra of the interacting galaxies
NGC~833 and NGC~835 by including the data of \Chandra \ and \XMM \ 
observed on multiple epochs.
Our main findings are summarized as follows.

\begin{itemize}

\item We have obtained the tightest upper limits of the hard X-ray flux
      above 10 keV for NGC~838 and NGC~839, supporting the previous 
      arguments by \citet{sul14} that they are both starburst dominant galaxies
      without luminous AGNs ($L_\mathrm{2-10}<3 \times 10^{42}$ erg s$^{-1}$
      even assuming a column
      density of $\log N_\mathrm{H} = 25$). 

\item We have confirmed that both NGC~833 and NGC~835 contain obscured
      LLAGNs with intrinsic 2--10 keV luminosities of $\approx 3 \times
      10^{41}$ erg s$^{-1}$ and line-of-sight column densities of
      $N_\mathrm{H}^\mathrm{LS} \approx 3\times10^{23}$
      cm$^{-2}$. Reprocessed X-ray radiation from cold matter is detected,
      indicating that the tori are moderately developed in both objects.

\item We have revealed that NGC~835 underwent long-term variability in
      both intrinsic luminosity and absorption. The line-of-sight column
      density was changed from $\approx 5 \times 10^{23}$~cm$^{-2}$ in
      2000 to $ \approx 3\times 10^{23}$~cm$^{-2}$ in 2013/2015. This
      can be interpreted as a transit of clouds as expected for
      clumpy tori.

\item We point out that there is large diversity in the relation between
      the 2--10 keV AGN luminosity and the 8--1000 $\mu$m infrared
      luminosity in local LLAGNs, regardless of their environments (in
      interacting systems or not). This is consistent with the general
      idea that the mass accretion process in the nucleus and the star
      forming activity in the disk are not strongly coupled.

\end{itemize}

\bigskip

\acknowledgements

Part of this work was financially supported by the Grant-in-Aid for
Scientific Research 17K05384 (Y.U.) and for JSPS Fellows for young
researchers (A.T.). We acknowledge financial support from the
China-CONICYT fellowship (C.R.), FONDECYT 1141218 (C.R.), and Basal-CATA
PFB-06/2007 (C.R.). This research has made use of the \NuSTAR \ Data
Analysis Software (\textsc{NuSTARDAS}) jointly developed by the ASI
Science Data Center (ASDC, Italy and the California Institute of
Technology (Caltech, USA).  Also this research has made use of the
NASA/IPAC Infrared Science Archive, which is operated by the Jet
Propulsion Laboratory, California Institute of Technology, under
contract with the National Aeronautics and Space Administration.

\end{document}